\documentclass[conference]{IEEEtran}
\IEEEoverridecommandlockouts

\usepackage{cite}
\usepackage{amsmath,amssymb,amsfonts}
\usepackage{algorithmic}
\usepackage{graphicx}
\usepackage{textcomp}
\usepackage{xcolor}

\usepackage{hyperref}
\usepackage[hyphenbreaks]{breakurl}
\usepackage{multirow}
\usepackage{subfigure}
\usepackage{algorithm}
\usepackage{tabularx}
\usepackage{color}
\usepackage{textcomp}

\usepackage{caption2} 

\usepackage{tikz}

\usepackage{filecontents}

\newcommand{\tabincell}[2]{\begin{tabular}{@{}#1@{}}#2\end{tabular}}
\begin{document}
\bstctlcite{IEEEexample:BSTcontrol}

\title{Consistent RDMA-Friendly Hashing on Remote Persistent Memory}
\author{
{   Xinxin Liu,
    Yu Hua,
    Rong Bai
}\\
Huazhong University of Science and Technology\\
}
\maketitle

\begin{abstract}
  Coalescing RDMA and Persistent Memory (PM) delivers high end-to-end performance for networked storage systems, which requires rethinking the design of efficient hash structures. In general, existing hashing schemes \emph{separately} optimize RDMA and PM, thus partially addressing the problems of \emph{RDMA Access Amplification} and \emph{High-Overhead PM Consistency}. In order to address these problems, we propose a continuity hashing, which is a ``one-stone-two-birds'' design to optimize both RDMA and PM. The continuity hashing leverages a fine-grained contiguous shared region, called SBuckets, to provide standby positions for the neighbouring two buckets in case of hash collisions. In the continuity hashing, remote read only needs a single RDMA read to directly fetch the home bucket and the neighbouring SBuckets, which contain all the positions of maintaining a key-value item, thus alleviating RDMA access amplification. Continuity hashing further leverages indicators that can be atomically modified to support log-free PM consistency for all the write operations. Evaluation results demonstrate that compared with state-of-the-art schemes, continuity hashing achieves high throughput (i.e., $1.45X$ -- $2.43X$ improvement), low latency (about $1.7X$ speedup) and the smallest number of PM writes with various workloads, while has acceptable load factors of about $70\%$.
\end{abstract}

\section{Introduction} \label{sec:introduction}
High-speed networks and efficient persistent storage contribute to the high performance of cloud applications. Remote direct memory access (RDMA) is able to support bypassing kernel to directly access a remote memory~\cite{RDMAAwareProgrammingusermanual}. RDMA enables zero memory copy and provides extremely high bandwidth and low latency. At the same time, the persistent memories (PMs), such as $3$D XPoint~\cite{3dxpoint} and PCM~\cite{wong2010phase}, blur the line between memory and storage, thus offering non-volatility, byte-addressability, large capacity and DRAM-scale latency. Therefore, many schemes coalesce RDMA and PMs to deliver high end-to-end performance~\cite{anderson2020assise,yang2020filemr,tsai2020disaggregating,yang2019orion,chen2019scalable,ma2020asymnvm,kim2018hyperloop,hu2018persistence,shan2017distributed,lu2017octopus,islam2016high,zhang2015mojim}.

The coalesced RDMA and PM require rethinking the design of high-performance index structures, which is important for efficient large-scale storage systems. In this paper, we focus on the hash-based index structures that enable fast point query response and have been widely used in networked databases and key-value stores~\cite{Memcached,Redis,Aerospike,Oracle}. However, applying hashing schemes to RDMA and PM environments needs to address two main challenges:

\textbf{RDMA Access Amplification.} RDMA is efficient and well-known for one-sided operations (e.g., RDMA read, RDMA write and atomic operations), which are able to thoroughly bypass remote CPUs and provide higher bandwidth and lower latency than two-sided operations over RC (reliable connection) mode~\cite{wei2018deconstructing}. However, a single one-sided RDMA operation only reads/writes one contiguous memory region. Therefore, accessing non-contiguous remote memory (e.g., following pointers in the chained hashing~\cite{knuth1998art} and querying data in $n$-way cuckoo hashing~\cite{kutzelnigg2008random,pagh2004cuckoo}) require multiple one-sided RDMA round-trips. We refer to this problem of high network overheads as \emph{RDMA Access Amplification}.

\textbf{High-Overhead PM Consistency.} Due to the existence of volatile parts in PM-based systems (e.g., the CPU caches), in order to ensure crash consistency in case of a system failure, updating data larger than the $8$-byte atomic write unit usually requires undo/redo logging or copy-on-write (COW)~\cite{shan2017distributed,nguyen2018picl,ogleari2018steal,nam2019write,dulloor2014system}. However, double write operations in these mechanisms consume the limited endurance of PM. Moreover, guaranteeing write ordering typically needs the aid of cache line flush (e.g., clflush, clflushopt and clwb) and memory barrier instructions (e.g., mfence and sfence)~\cite{IntelArchitectureInstruction}, thus resulting in high system performance overheads~\cite{chen2015persistent,venkataraman2011consistent}.

Existing hashing schemes separately optimize RDMA or PM, and partially address the above challenges. Specifically, \textbf{RDMA-friendly hashing schemes} are usually designed to address the problem of \emph{RDMA access amplification}~\cite{dragojevic2014farm,dragojevic2015no,shamis2019fast,wei2015fast,chen2016fast,wei2018deconstructing}. FaRM~\cite{dragojevic2014farm,dragojevic2015no,shamis2019fast} proposes a hopscotch hashing-based algorithm that combines chaining and associativity, enabling a small number of one-sided RDMA reads for remote lookups. DrTM~\cite{wei2015fast,chen2016fast,wei2018deconstructing} proposes a HTM/RDMA-friendly hashing, called cluster hashing. The cluster hashing leverages clustering (i.e., clustering multiple keys into a bucket) and location-based caches to reduce the number of RDMA operations. However, these RDMA-friendly hashing schemes fail to mitigate \emph{High-Overhead PM Consistency}. For \textbf{PM-friendly hashing schemes}, many works have been proposed to guarantee crash consistency and optimize PM writes~\cite{nam2019write,zuo2018write,lee2019recipe}. Level hashing~\cite{zuo2018write} is a write-optimized two-level (i.e., the top level and the bottom level) hashing scheme for PM. It leverages the tokens contained in each bucket and the two-level structure to provide cost-efficient resizing and low-overhead consistency guarantee. Cacheline-conscious extendible hashing (CCEH)~\cite{nam2019write} is a PM-friendly three-level (i.e., a global directory, segments and cache-line sized buckets) dynamic hashing, which guarantees crash consistency without explicit logging. RECIPE~\cite{lee2019recipe} proposes a general method for converting DRAM-based indexes that meet the specified conditions into PM-based indexes with crash consistency guarantees. However, these PM-friendly hashing schemes typically cause \emph{RDMA Access Amplification} due to indirect layers~\cite{nam2019write} or non-contiguous standby positions~\cite{zuo2018write}.

Unlike existing hashing schemes, we propose a coalescing hashing solution for both RDMA and PM, called continuity hashing, which mitigates \emph{RDMA access amplification} and PM writes, as well as guaranteeing PM crash consistency. Specifically, this paper makes the following contributions:

\textbf{Coalescing Design for RDMA and PM.} We propose the continuity hashing index structure, a ``one-stone-two-birds'' design to optimize both RDMA and PM, i.e., alleviating \emph{RDMA access amplification} and providing consistency guarantee with optimized PM writes. In the continuity hashing, two buckets with adjacent bucket numbers share a fine-grained contiguous memory region, called shared buckets (SBuckets). These SBuckets provide standby positions for the neighbouring two buckets in case of hash collisions. Clients use one-sided RDMA reads to complete read requests for saving the server's CPUs and delivering high performance, and write requests are handled by the server in order to simplify read-write competition and ensure consistency with low overheads.

\textbf{Remote Read without Access Amplification.} In our continuity hashing, a bucket and the neighbouring SBuckets build a small contiguous memory region, called a segment, which contains all the potential positions of a specific key-value item. Therefore, to obtain a requested record, clients compute the remote location based on the key's hash value, and only need a single RDMA read to directly fetch the corresponding segment, thus reducing the potential multiple RDMA round-trips.

\textbf{Log-Free Consistency Guarantee.} Two segments in the continuity hashing overlap on the neighbouring SBuckets. We use an indicator for each two overlapping segments to indicate whether each slot in the two segments contains a consistent key-value item. An indicator can be modified with an $8$-byte atomic write, thus supporting log-free consistency guarantee for all the write operations on PM. Moreover, since an indicator locates at the beginning of the SBuckets that belong to the two overlapping segments, it can be remotely fetched together with a segment using one RDMA read operation, and guide clients to obtain consistent data.

We implemented the continuity hashing that is compared with state-of-the-art schemes, including level hashing~\cite{zuo2018write} (i.e., PM-friendly hashing scheme) and P-FaRM-KV (i.e., a PM implementation of RDMA-friendly hashing scheme FaRM-KV~\cite{dragojevic2014farm} converted via RECIPE~\cite{lee2019recipe}). Compared with them, extensive evaluation results demonstrate that our continuity hashing achieves the highest throughputs (i.e., $1.45X$ -- $2.43X$) with various workloads. For latencies, the continuity hashing has better search performance than P-FaRM-KV, and significantly outperforms the level hashing by an average of $2.19X$. The continuity hashing also has better write performance than level hashing, and further achieves a $1.99X$ improvement on average compared with P-FaRM-KV. Continuity hashing achieves the smallest number of PM writes, and obtains acceptable load factors of about $70\%$ given the large capacity of the available PM products~\cite{izraelevitz2019basic}.

The rest of this paper is organized as follows. Section~\ref{sec:background} presents the background and motivation. Section~\ref{sec:design} describes
the design of continuity hashing and Section~\ref{sec:implementation} presents the implementation. The performance evaluation is shown in Section~\ref{sec:evaluation}. Section~\ref{sec:related work} discusses the related work and Section~\ref{sec:conclusion} concludes this paper.

\section{Background and Motivation}\label{sec:background}

\subsection{Remote Direct Memory Access}
\begin{figure}[!ht]
	\vspace{-0.2cm}
	\centering
	\setlength{\abovecaptionskip}{0.15cm}	
	\includegraphics[width=0.46\textwidth]{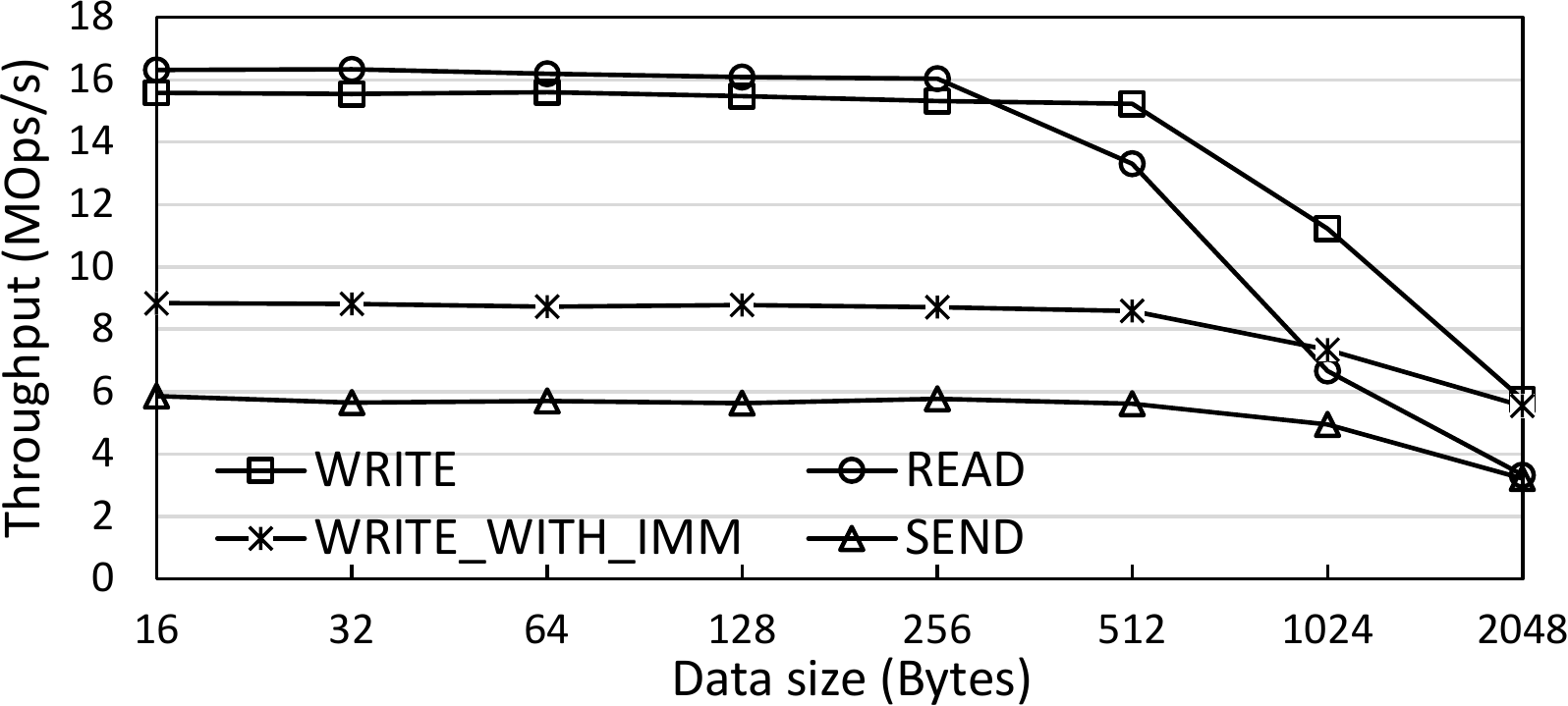}
	\caption{\label{fig:primitive-performance}A comparison of different RDMA primitives.}
	\vspace{-0.5cm}
\end{figure}

Remote direct memory access (RDMA) enables to bypass kernels to directly access a remote memory, thus supporting zero memory copy and providing extremely low latency and high bandwidth~\cite{novakovic2019storm,wei2018deconstructing,kalia2016design,tsai2017lite,kalia2014using}. RDMA has two kinds of operations, i.e., two-sided and one-sided. Two-sided operations, such as RDMA send and recv, are served by the remote CPU, which needs to poll RDMA messages and process them~\cite{RDMAAwareProgrammingusermanual}. Two-sided operations are similar to socket programming~\cite{RDMAAwareProgrammingusermanual}, but the remote procedure call (RPC) using two-sided RDMA operations introduces much higher throughput than RPC using TCP/IP due to RDMA's ability to bypass kernel and traditional network stacks~\cite{wei2018deconstructing,kalia2016fasst,dragojevic2014farm}. In general, the request/reply pattern of two-sided operations involves the communications of two underlying RDMA verbs. Unlike two-sided operations, one-sided operations include RDMA read, RDMA write and atomic operations (Fetch and Add/Compare and Swap). When accessing the remote memory, the one-sided operations do not involve remote CPU and allow one machine to directly access the remote memory of another machine without any prior knowledge of the remote process. Furthermore, one-sided operations have higher throughput and lower latency than two-sided counterparts~\cite{mitchell2013using,wei2018deconstructing,lu2017octopus,dragojevic2014farm}. Because one-sided operations have less bookkeeping than two-sided ones. Figure~\ref{fig:primitive-performance} shows a primitive-level comparison on two Linux machines with $100$Gbps Infiniband network and $256$GB Optane DIMMs. The results show that one-sided primitives outperform two-sided primitives for the data whose sizes are smaller than $2,048$B.

RDMA exhibits three transport modes, i.e., Reliable Connection (RC), Unreliable Connection (UC) and Unreliable Datagram (UD)~\cite{RDMAAwareProgrammingusermanual}. Specifically, RC offers reliable data transmissions and the delivered packets become in order. The connection in UC mode is not reliable, and a higher level protocol needs to handle errors, e.g., packet loss. The UD mode does not guarantee the reliable delivery and  accurate ordering, and the receivers possibly drop packets. Different transport modes support different RDMA operations. UD only supports two-sided operations (i.e., RDMA send and recv), and UC supports RDMA send/recv and RDMA write. RC supports both one-sided and two-sided operations (i.e., RDMA send/recv, RDMA read/write and atomic operations). In this paper, we use RC transport mode to support both one-sided and two-sided RDMA operations, as well as eliminate the extra complexity from upper protocols or applications to guarantee the delivery and ordering.

\subsection{Persistent Memory}
Persistent Memory (PM) devices have the strengths of non-volatility, byte-addressability, large capacity and DRAM-scale latency, which blur the boundary between memory and storage.

However, due to suffering from limited write endurance~\cite{zhou2009durable,yang2013memristive,qureshi2009enhancing}, PM-based systems need to efficiently address the persistence and consistency problems in case of power failure or system crash~\cite{liu2018crash}. Specifically, RDMA NIC (RNIC) fails to guarantee the persistence with PMs~\cite{yang2019orion,kim2018hyperloop} because the completion of RDMA writes from the client's view merely means that the delivered data have reached the remote RNIC, which possibly fail to be persistent in PMs. Intel hence proposed that one can issue an extra RDMA read or send command after RDMA write(s) to guarantee the persistence for one-sided RDMA writes~\cite{ChetDouglasRDMAwithPM,ChetDouglasIntelPerspective}, however resulting in high network overheads. Therefore, we prefer to leverage the CPU in a server to ensure data persistence. Moreover, PM systems typically contain volatile CPU caches. Executing the in-place updating upon data, whose size is larger than the failure atomicity unit (i.e., $8$ bytes), possibly results in data corruption, if a system failure occurs. Undo logging, redo logging and copy-on-write (COW) are employed to address the above consistency problem~\cite{shan2017distributed,nguyen2018picl,ogleari2018steal,nam2019write,dulloor2014system}. Specifically, the undo/redo logging needs to first append old/new data into the undo/redo logs and then perform updates, while COW needs to first create a data copy and then perform update operations on the copy, resulting in double PM writes. Moreover, since CPU reorders the two-time memory writes, programmers need to use cache line flush instructions (e.g., clflush, clflushopt and clwb) and memory barrier instructions (e.g., mfence and sfence) to guarantee write ordering and data consistency~\cite{IntelArchitectureInstruction}. These cache line flush and memory barrier instructions result in high system performance overheads~\cite{chen2015persistent,venkataraman2011consistent}. In summary, it is important to maintain crash consistency and reduce PM writes in persistent memory systems.

\subsection{Hashing Indexes for RDMA and PM}
\subsubsection{RDMA-friendly Hashing Schemes}
In order to efficiently leverage the one-sided RDMA operations that support high bandwidth, low latency and bypassing remote CPUs, existing RDMA-friendly hashing schemes usually collect multiple key-value pairs with the same hash value (i.e., items with hash collisions) in a small contiguous region~\cite{wei2015fast,dragojevic2014farm}. The contiguous memory region can be retrieved with a single RDMA read, instead of multiple one-sided RDMA operations, thus reducing network overheads.

DrTM-KV~\cite{wei2015fast} proposes a cluster chaining hash table, called cluster hashing, to exploit RDMA and HTM (hardware transactional memory) transactions. Cluster hashing is based on the traditional chained hashing where linked lists are used to deal with hash collisions. To reduce the length of the linked lists and reduce the number of one-sided RDMA operations, DRTM-KV leverages a clustering technique, i.e., clustering multiple keys (such as $8$) into a bucket. DrTM-KV also builds a location-based cache that caches the location of key-value pairs to further reduce the number of RDMA operations.

FaRM-KV~\cite{dragojevic2014farm} proposes chained associative hopscotch hashing that combines hopscotch hashing~\cite{HopscotchHashing} with chaining and associativity. In hopscotch hashing, a key-value pair is stored in the bucket that the key is hashed to (i.e., the home bucket) and the following $H-1$ buckets (i.e., the neighbourhood of the home bucket). These $H$ buckets locate in a contiguous memory region and can be fetched using a single RDMA operation. To insert a key-value item, hopscotch hashing first looks for an empty bucket within the neighbourhood via linear probing~\cite{pittel1987linear}. If all the $H$ buckets are full, hopscotch hashing moves an empty bucket into the neighbourhood by iteratively displacing existing key-value items. However, if hopscotch hashing still cannot find an empty bucket to store the new key-value item, the hash table will be resized. FaRM-KV adds associativity (i.e., $H/2$ slots per bucket) and chaining to the hopscotch hashing. If no empty bucket can be moved into the neighbourhood, FaRM-KV will add an overflow chain to the home bucket, rather than resizing the hash table.

However, directly applying these RDMA-friendly hashing schemes to a PM platform without battery backup will result in the crash consistency problem due to the volatile parts of PM systems (e.g., caches). In addition, RDMA-friendly hashing schemes generally lack an optimized design for PM write operations, exacerbating the limited endurance of PM.

\subsubsection{PM-Friendly Hashing Schemes}
In order to exploit the non-volatility and large capacity properties of PM and improve index efficiency and reliability, a number of PM-friendly hashing schemes~\cite{nam2019write,zuo2018write,lee2019recipe} have been proposed. Compared with DRAM-based hashing, PM-based hashing needs to ensure crash consistency in case of power failure or system crash via atomic write, logging or COW techniques with the aid of cache line flush and mfence instructions. In addition, PM-friendly hashing schemes generally optimize PM writes to improve system performance and PM endurance.

Cacheline-conscious extendible hashing (CCEH)~\cite{nam2019write} is a PM-friendly variant of extendible hashing~\cite{fagin1979extendible} that dynamically expands and shrinks on demand. CCEH structure contains three levels, i.e., a global directory that stores segment addresses, segments each containing a group of buckets and cache-line sized buckets. For a given hash value of a key-value item, CCEH leverages the first several leading bits (such as $2$) as the segment index and the least significant byte as the bucket index. Each query needs to access the directory entry using the segment index to obtain the address of the corresponding segment, and then using the segment address and the bucket index to access the corresponding bucket that contains the requested record in the segment. In addition, CCEH guarantees crash consistency via enforcing the ordering of PM writes.

Level hashing~\cite{zuo2018write} is a write-optimized hashing scheme for PM. It has two levels, namely, the top level and the bottom level, and the capacity of the former is twice that of the latter. Level hashing uses two independent hash functions, and each bucket in the bottom level provides a standby position for two buckets in the top level to deal with hash collisions. Therefore, a given key-value item is able to be stored in four separate buckets. During resizing, level hashing rehashes the items in the original bottom level to the newly allocated top level whose size is $4$ times than the size of the original bottom level, and the original top level becomes the current bottom level. To provide low-overhead consistency guarantee, level hashing leverages the tokens contained in each bucket to provide log-free deletion, insertion and resizing as well as opportunistic log-free update. If there is no empty slot in the bucket where the key-value item to be updated is located, the level hashing will use the logs to ensure consistency.

However, directly applying these PM-friendly hashing schemes to RDMA environments generally results in \emph{RDMA Access Amplification}, which means one has to issue multiple one-sided RDMA operations, instead of a single one, due to the non-continuity of memory regions to be accessed. For example, each remote read in CCEH requires two one-sided RDMA operations due to the indirection introduced by the directory, while in level hashing, querying a key-value item needs to access at most four non-contiguous buckets with four one-sided RDMA reads, resulting in high network overheads.

\begin{figure*}[!ht]
	\centering
	\setlength{\abovecaptionskip}{0.2cm}	
	\includegraphics[width=0.96\textwidth]{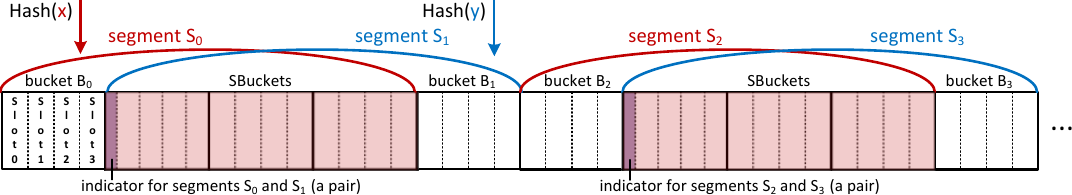}
	\caption{\label{fig:Hash-Index-Structure}Continuity hashing index structure.}
	\vspace{-0.3cm}
\end{figure*}

Unlike existing hashing schemes that separately optimize RDMA or PM, we aim to propose a coalescing hashing solution for both RDMA and PM.

\section{The Design of Continuity Hashing}\label{sec:design}
In this Section, we present the design of continuity hashing and the design goals aim to reduce RDMA access amplification, maintain PM crash consistency and reduce PM writes. We further present the index structure of continuity hashing, the read/write operations using RDMA, and the log-free failure-atomicity guarantee.

\subsection{Continuity Hashing Index Structure}\label{subsec:Hash Table Structure}
Continuity hash tables have two kinds of hash buckets, i.e., the numbered buckets and the unnumbered shared buckets (SBuckets). Specifically, the numbered buckets are addressable by a hash function, and two numbered buckets with adjacent bucket numbers (but non-contiguous memory addresses) share a fine-grained contiguous memory region between them, which consist of several (e.g., $3$) unnumbered SBuckets. For example, as shown in Figure~\ref{fig:Hash-Index-Structure}, buckets $B_0$ and $B_1$ share three unnumbered SBuckets, and buckets $B_2$ and $B_3$ also share three unnumbered SBuckets. These SBuckets store conflicting key-value items when the corresponding home buckets (i.e., the neighbouring two numbered buckets) are full, playing a similar role of additional storage named stash~\cite{debnath2016PFHT,kirsch2010more}. But the differences are that 1) we organize all the numbered buckets and the unnumbered SBuckets into a contiguous memory region, and 2) we adopt a fine-grained shared region between each two numbered buckets. The design of the contiguous memory region for hash collisions comes from the RDMA property that each one-sided RDMA operation can only access one contiguous memory region. Therefore, all the potential positions of a specific key-value item are in a small contiguous memory region, which can be fetched with a single one-sided RDMA operation, thus reducing potential multiple RDMA round-trips. The fine-grained shared region is  further designed as a suitable trade-off between the high space utilization of the hash table and the small size of the data retrieved by the client via one-sided RDMA. Specifically, extending the shared regions can mitigate hash collisions and thus improve the space utilization of the hash table.

A large shared region (e.g., a stash for the entire hash table) increases the payload of RDMA read, thus reducing system throughput. Therefore, we adopt the design decision that pairs of adjacent numbered buckets share a fine-grained shared region, and the number of SBuckets in the fine-grained shared regions can be flexibly configured each time resizing occurs. Considering that small key-value pairs dominate in real-world key-value workloads~\cite{lai2015atlas,atikoglu2012workload}, in order to improve cache efficiency, we enable each bucket to contain multiple slots like CCEH~\cite{nam2019write}, level hashing~\cite{zuo2018write}, PFHT~\cite{debnath2016PFHT} and bucketized cuckoo hashing~\cite{breslow2016horton,li2014algorithmic,fan2013memc3}.

As the basic unit of a one-sided RDMA read operation from clients, a segment is interpreted as a numbered bucket and a neighbouring fine-grained shared region (i.e., several unnumbered SBuckets). As shown in Figure~\ref{fig:Hash-Index-Structure}, segment $S_0$ consists of bucket $B_0$ and the subsequent three unnumbered SBuckets that are shared by buckets $B_0$ and $B_1$, while segment $S_3$ consists of bucket $B_3$ and the preceding three unnumbered SBuckets that are shared by buckets $B_2$ and $B_3$. Each two segments with adjacent segment numbers overlap on a fine-grained shared region. For example, in Figure~\ref{fig:Hash-Index-Structure}, segments $S_0$ and $S_1$ overlap on three unnumbered SBuckets between buckets $B_0$ and $B_1$, and segments $S_2$ and $S_3$ overlap on three unnumbered SBuckets between buckets $B_2$ and $B_3$. For the convenience of description, we call each two overlapping segments (e.g., segments $S_0$ and $S_1$) a \emph{segment pair}.

We use an indicator for each segment pair to indicate whether each slot in the two segments contains valid data. Specifically, the number of bits per indicator is the slot number in the corresponding segment pair, and each indicator is stored at the beginning of the fine-grained shared region that belongs to the segment pair. Therefore, when a client issues an RDMA read to fetch a segment, the corresponding indicator indicates the slots with the valid data, without the need for another network round-trip. As shown in Figure~\ref{fig:Hash-Index-Structure}, a segment pair has two numbered buckets and three unnumbered SBuckets. The sum of slots in the two segments is $5 * 4 = 20$. Therefore, a $20$-bit indicator for a segment pair is sufficient, which can be modified with an $8$-byte atomic write. In general, an indicator for each segment pair has three advantages. First, the indicator helps ensure crash consistency without expensive logging. Second, the indicator reduces the number of cache line accesses by enabling the server to directly locate an empty slot when processing write requests. Third, the indicator is at the beginning of the unnumbered SBuckets and can be fetched together with a segment by clients via one RDMA read operation. The details are described in Sections~\ref{subsec:RDMA Procedures} and~\ref{subsec:Consistency Guarantee}.

\subsection{Read/Write Operations using RDMA}\label{subsec:RDMA Procedures}
The communications between the server and clients leverage the fast RDMA networking. To allow RDMA operations, when clients first establish the connections to a server, the server registers the memory regions of the continuity hash table with RNIC (RDMA-enabled NIC), and clients obtain the corresponding remote registration keys, as well as the address and the size of the remote hash table. Clients further issue RDMA operations to these registered memory regions.

For the procedure of remote reads from clients to the server, we explore the advantages of one-sided RDMA operations that do not involve remote server's CPU and have higher bandwidth and lower latency than two-sided RDMA operations~\cite{mitchell2013using,wei2018deconstructing,lu2017octopus,dragojevic2014farm}. Clients compute the remote location (i.e., the bucket number) of a requested key based on its hash value:
\begin{equation}\label{BucketNumber}
   Bucket\_Number = hash(k)\%N
\end{equation}
where $hash(k)$ is the hash value of the requested key, and $N$ is the total number of the numbered buckets. Considering that a one-sided RDMA read accesses at most $1$GB contiguous memory region~\cite{RDMAAwareProgrammingusermanual}. In order to reduce the number of network round-trips, clients only use one RDMA read to directly obtain the corresponding segment that contains all the potential locations of the requested key-value item in the continuity hash table. If the computed bucket number is even (e.g., bucket $B_0$), the offset of the segment to be read remotely (e.g., segment $S_0$) is:
\begin{equation}\label{BucketNumber}
   Ofs = hash(k) \% N / 2 * (size_{se} + size_{bu})
\end{equation}
If the computed bucket number is odd (e.g., bucket $B_1$), the offset of the segment to be read remotely (e.g., segment $S_1$) is:
\begin{equation}\label{BucketNumber}
   Ofs = (hash(k)\%N - 1)/2 * (size_{se} +size_{bu})+ size_{bu}
\end{equation}
where $size_{se}$ and $size_{bu}$ are the segment size and the bucket size respectively. Subsequently, if the requested key exists in the continuity hash table, clients will find the requested key-value item locally with the aid of the indicator. The procedure of remote reads via one-sided RDMA finishes.

For the procedure of write requests from clients to the server, instead of using the one-sided RDMA write operations that bypass the remote CPU, we put the burden of handling write requests on the server. This design choice is based on three reasons. \emph{First}, directly writing to a continuity hash table by multiple clients via one-sided RDMA writes requires locking over the network (e.g., using atomic RDMA compare-and-swap operations), which leads to high network overheads. In addition, we also need to use additional mechanisms to handle the cases where a client holding a lock fails and other clients are kept waiting to acquire the lock. \emph{Second}, without the coordination between the local memory accesses initiated by the server's CPU and the remote memory access using RDMA~\cite{mitchell2013using}, RDMA atomic operations (e.g., compare-and-swap) don't provide atomicity with respect to the remote CPU's atomic operations. Hence, one-sided RDMA writes from clients are not aware of the local locks on the server, thus causing data inconsistency. \emph{Third}, existing RDMA NICs fail to ensure the persistence for one-sided RDMA writes to persistent memory~\cite{yang2019orion,kim2018hyperloop}. A general method to guarantee persistence and consistency for one-sided RDMA writes is to issue an extra RDMA read or an extra RDMA send after RDMA write(s)~\cite{ChetDouglasRDMAwithPM,ChetDouglasIntelPerspective}, resulting in extra network round-trips. We prefer to use server's CPU to guarantee persistence, like existing RDMA-based persistent memory systems~\cite{yang2019orion,shan2017distributed,zhang2015mojim}.

In summary, for the procedure of remote write requests, clients query the server using RDMA write\_with\_immediate operations, which are available in the RDMA communication and similar to one-sided RDMA writes, except notifying the remote server of the immediate values~\cite{RDMAAwareProgrammingusermanual}. RDMA write\_with\_immediate operations have higher throughput than RDMA send/recv operations over RC (Reliable Connection)~\cite{wei2018deconstructing}, as shown in Figure~\ref{fig:primitive-performance}. In our design, we place the client's identifier into the immediate data field. After receiving remote write requests from clients, the server processes these requests and then notifies the clients that their requests have been completed.

\subsection{Log-Free Failure-Atomicity Guarantee}\label{subsec:Consistency Guarantee}
In this Section, we introduce the local writing process of the server with log-free failure-atomicity guarantee. Existing hash tables use a $1$-bit token that is associated with a slot to indicate whether the corresponding slot is empty~\cite{zuo2018write,zuo2017write,fan2013memc3}. We adopt this design and group a set of tokens, called an indicator for each segment pair. As illustrated in Section~\ref{subsec:Hash Table Structure}, an indicator is able to be updated in the atomic-write manner and, enables a log-free failure-atomicity guarantee for all the write operations (i.e., insertion, deletion, update and resizing) on persistent memory. The mfence instruction needs to guarantee the order requirements in the following operations.

\emph{Atomic Insertion: } To insert a new key-value item, our continuity hashing computes the home location (i.e., the home bucket number) of the requested item via Equation~\eqref{BucketNumber}. An even bucket number (e.g., $B_0$) means that the corresponding fine-grained shared region, or the unnumbered SBuckets, is on the right side of the home bucket. Moreover, the server sequentially checks each bit in the indicator corresponding to the home bucket and the SBuckets from left to right. For example, in Figure~\ref{fig:Hash-Index-Structure}, if the home bucket is bucket $B_0$ , the server will sequentially checks the first $16$ bits of the indicator that are associated with each slot in bucket $B_0$ and the unnumbered SBuckets (i.e., segment $S_0$). Once a bit in the indicator is found to be equal to $0$, which means the associated slot in the home bucket or the SBuckets is empty, the server will stop checking the remaining bits in the indicator and inserts the requested item into the associated empty slot. Unlike it, an odd bucket number (e.g., $B_1$) means that the corresponding unnumbered SBuckets is to the left of the home bucket. The server checks each associated bit in the indicator in the reverse order, and inserts the requested item into the first empty slot. Finally, after the requested key-value item is written to the empty slot, the server atomically sets the associated bit in the indicator from $0$ to $1$, and the insert operation completes successfully. Even if a power failure or system crash occurs during writing the key-value pair, the continuity hash table is still in a consistent state. Because the associated bit in the indicator has not been changed and thus the partial write is not visible.

\emph{Atomic Deletion: } To delete a key-value pair, the server computes the home bucket number and then queries the corresponding segment (i.e., the home bucket and the unnumbered SBuckets) by using the indicator. The direction of the query in the segment is based on the parity of the computed bucket number like that in the insert operations. After finding the key-value item to be deleted, the server only needs to set the associated bit in the indicator from $1$ to $0$ in the atomic-write manner, and the key-value item will be considered invalid by subsequent requests.

\emph{Atomic Update: } Continuity hashing adopts out-of-place update, which is a coalescence of insert and delete operations. Specifically, to update a key-value pair, the server locates the requested key-value pair in the continuity hash table like the delete operations. The server further attempts to identify an empty slot in the same segment, like the insert operation. Since the old and the new locations of the requested key-value pair are associated with the same indicator in the same segment, the server changes the values of the two corresponding bits in the indicator with an $8$-byte atomic write. The update to the key-value pair is invisible until the atomic update in the indicator is completed, thus ensuring the consistency of the continuity hash table even in the case of a system failure.

\emph{Log-free Resizing: } Resizing requires rehashing existing key-value items into a new hash table. Specifically, the server reads a key-value pair in the original hash table, then inserts it to the new hash table, and finally deletes the item from the original hash table. After all the items in the original hash table are rehashed in order, the server reclaims the memory space of the old hash table, and the resizing is completed. However, unlike the atomic update operations, the insertion and deletion for a key-value pair during resizing cannot be completed atomically due to the updates to two different indicators. In fact, the operation sequence (first insert and then delete an key-value item) ensures that the continuity hash table will not lose data in the event of a system failure. After restarting the server, we check the first existing key-value item of the old hash table and perform a delete or rehash operation based on whether it has been inserted into the new hash table, thus restoring the hash table to a consistent state.

\section{Implementation}\label{sec:implementation}
In this Section, we introduce how the continuity hashing supports concurrent read and write operations, and provide an optimization scheme for improving space utilization.

\subsection{Concurrent Reads and Writes}\label{subsec:Races}
In our design, write requests are processed by the server, and read requests are handled by clients. Hence, there exists the read-write competition between remote reads and local writes, which is able to guarantee the data consistency due to the atomic updates to indicators in the last step of write operations. For example, if a client uses a one-sided RDMA read to fetch a segment modified by a server, any partial key-value pairs that are being modified in the segment are invisible to the client since the server has not configured the indicator, thus ensuring consistency.

To support concurrent write operations initiated by different threads on the server, we allocate a spin-lock for each slot as well as its associated bit in an indicator. For the insertion and deletion for a key-value pair, a thread merely locks one slot as well as its associated bit before modifying them. For an update operation, a thread needs to lock two slots, i.e., the current slot and the target empty slot, before writing the new key-value pair into the empty slot and atomically updating the two associated bits in the indicator.

\subsection{Optimizing Space Utilization}\label{subsec:Space Utilization}
Our proposed continuity hashing aims to mitigate the \emph{access amplification} via one-sided RDMA. In the structure of continuity hashing, we use a single hash function by default. However, the load factors with only one hash function in traditional hash tables, where the load factor refers to the ratio of the number of stored KV items to that of total storage units, are pretty low. For example, in cuckoo hashing~\cite{CuckooHashing}, when the number of cells per hash bucket is $4$, the load factor with one hash function is merely $3\%$. When the number of cells per hash bucket reaches $8$, the load factor with one hash function is $12\%$~\cite{erlingsson2006cool}. Our design of the fine-grained shared region significantly improves the load factor with one hash function, though there is still room to improve the space utilization. In practice, not too high space utilization is acceptable due to the large capacity and the low price of the available PM products. For example, a single CPU is able to host $3$TB Intel Optane DC Persistent Memory~\cite{izraelevitz2019basic}, and the price of Optane is about half that of DRAM~\cite{IntelsOptaneDIMMPriceModel}. But we still provide an optimization for the space utilization or the load factor without violating our design goals.

\begin{figure}[!ht]
	\centering
	\setlength{\abovecaptionskip}{0.2cm}	
	\includegraphics[width=0.48\textwidth]{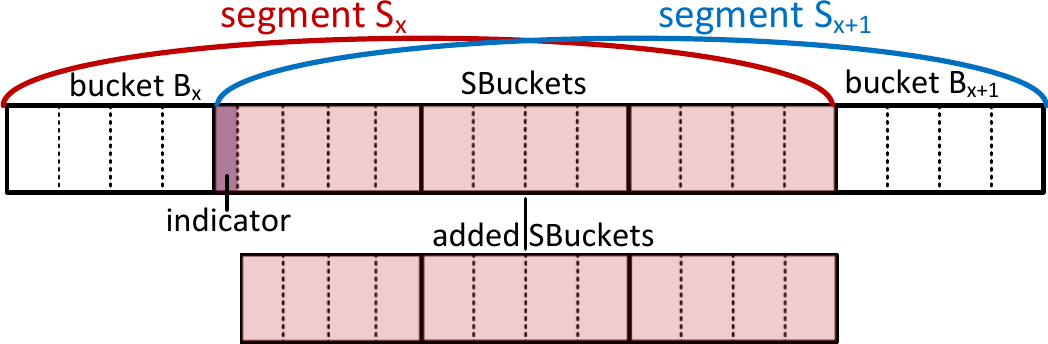}
	\caption{\label{fig:Added-SBuckets}Continuity hashing dynamically increases the number of SBuckets for $1/10$ segment pairs to improve load factors.}
\end{figure}

As shown in Figure~\ref{fig:Added-SBuckets}, to improve the load factor of continuity hashing, we add a new scheme that dynamically increases the number of SBuckets for a small percentage of segment pairs in the continuity hash table before the resizing operation. When the new SBuckets for a segment pair are added, the server will return the address and rkey of the added SBuckets region to the connected clients. The information (i.e., the address and rkey) is valid until the hash table is resized, and can be cached on clients. Clients can locally know whether a segment pair has added SBuckets. Specifically, each segment pair will have at most one SBucket group added before a resize operation is triggered, and we empirically set the percentage of segment pairs with added SBuckets to $1/10$ by running different configurations, which means continuity hashing can dynamically add SBuckets for at most $1/10$ segment pairs before resizing. In this case, for a uniform read workload, a client needs at most two RDMA round-trips (one for the segment and the other for the added SBuckets) with a $10\%$ probability, and only one RDMA round trip is required with the $90\%$ probability. Compared with the approach of using two hash functions where each remote read requires at most two RDMA round-trips, we propose to dynamically increase SBuckets, which incurs fewer network round-trips. In addition, our proposed method still supports log-free consistency for all write operations on persistent memory. Because the added SBuckets use the same indicator as the original buckets in the segment pair, and the indicator ($32$ bits in the example of Figure~\ref{fig:Added-SBuckets}) is able to be updated with an $8$-byte atomic write.

\begin{figure*}
	\begin{minipage}[t]{0.33\linewidth} 
		\centering
		\setlength{\abovecaptionskip}{0.1cm}
		\includegraphics[width=2.23in]{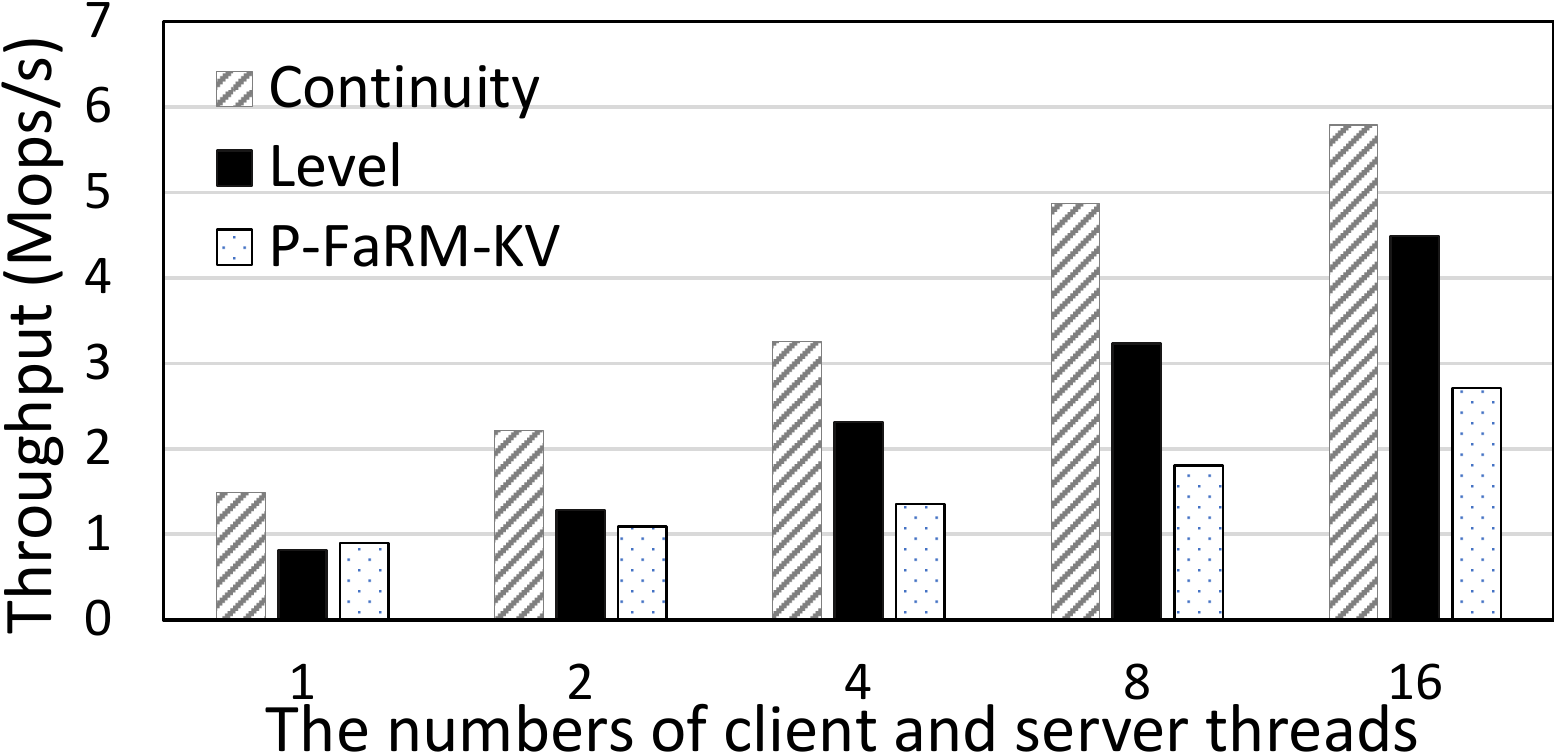}
		\setlength{\abovecaptionskip}{0.15cm}
		\setcaptionwidth{2.2in}
		\caption{\label{fig:throughput-workloada}The average throughput of YCSB-A (the update-heavy workload).}
	\end{minipage}
	\begin{minipage}[t]{0.33\linewidth}
		\centering
		\setlength{\abovecaptionskip}{0.1cm}
		\includegraphics[width=2.23in]{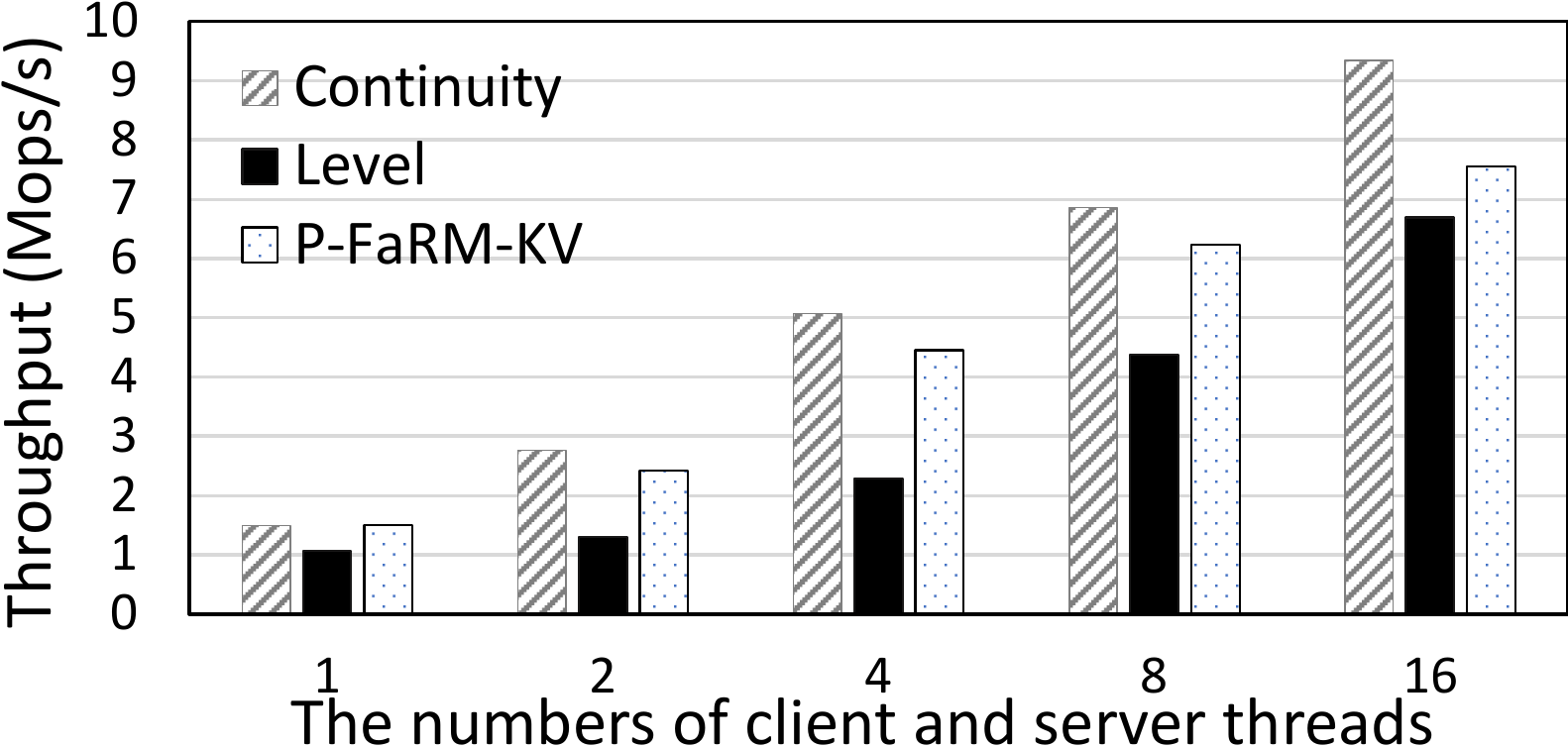}
		\setlength{\abovecaptionskip}{0.15cm}
		\setcaptionwidth{2.2in}
		\caption{\label{fig:throughput-workloadb}The average throughput of YCSB-B (the read-mostly workload).}
	\end{minipage}
	\begin{minipage}[t]{0.33\linewidth}
		\centering
		\setlength{\abovecaptionskip}{0.1cm}
		\includegraphics[width=2.23in]{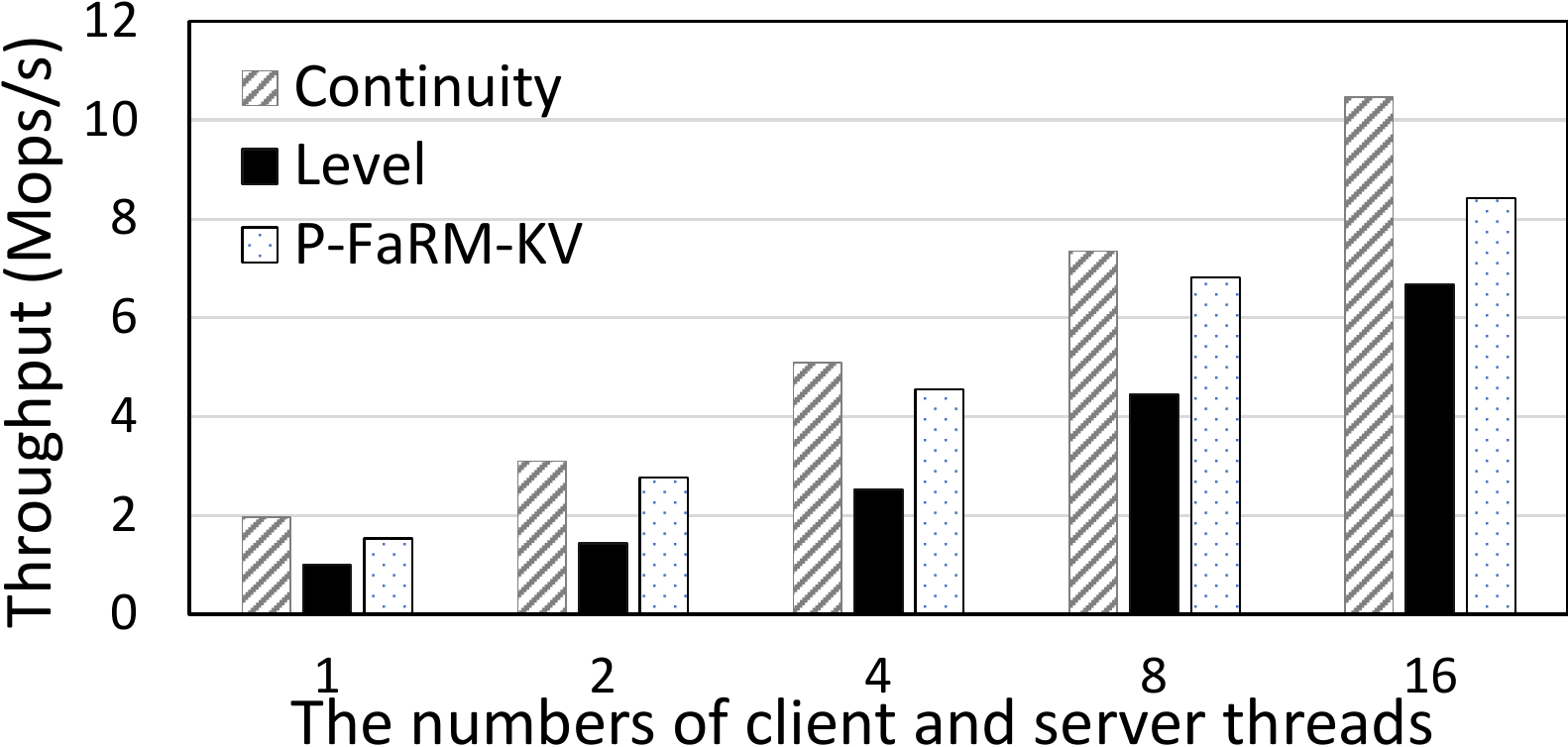}
		\setlength{\abovecaptionskip}{0.15cm}
		\setcaptionwidth{2.2in}
		\caption{\label{fig:throughput-workloadc}The throughput of YCSB-C (the read-only workload, positive searches).}
	\end{minipage}
	\vspace{-0.3cm}	
\end{figure*}

\begin{figure*}
	\begin{minipage}[t]{0.33\linewidth} 
		\centering
		\setlength{\abovecaptionskip}{0.1cm}
		\includegraphics[width=2.23in]{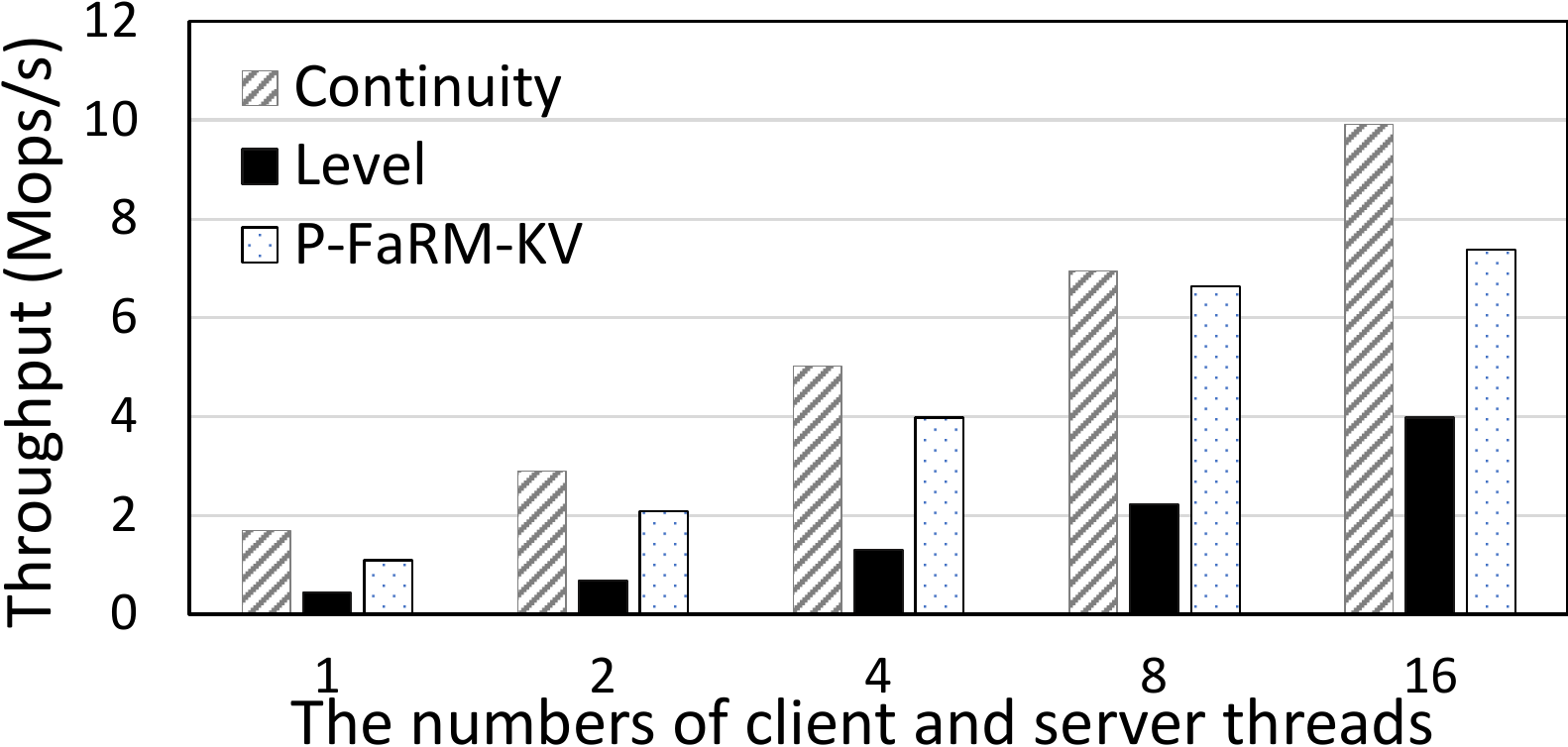}
		\setlength{\abovecaptionskip}{0.15cm}
		\setcaptionwidth{2.2in}
		\caption{\label{fig:throughput-workloadc-negative}The average throughput of negative searches.}
	\end{minipage}
	\begin{minipage}[t]{0.33\linewidth}
		\centering
		\setlength{\abovecaptionskip}{0.1cm}
		\includegraphics[width=2.23in]{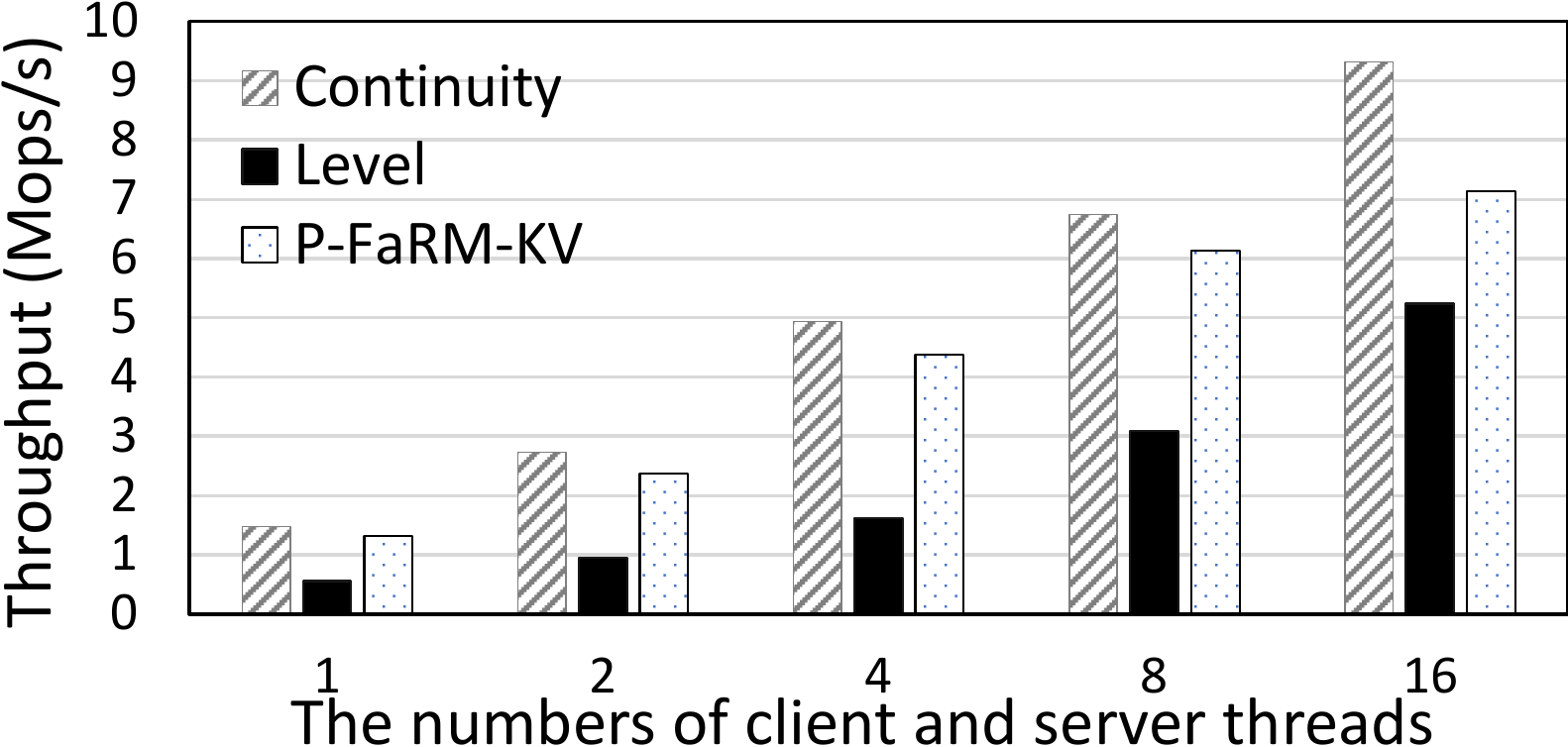}
		\setlength{\abovecaptionskip}{0.15cm}
		\setcaptionwidth{2.2in}
		\caption{\label{fig:throughput-workloadd}The average throughput of YCSB-D (the read-latest workload).}
	\end{minipage}
	\begin{minipage}[t]{0.33\linewidth}
		\centering
		\setlength{\abovecaptionskip}{0.1cm}
		\includegraphics[width=2.23in]{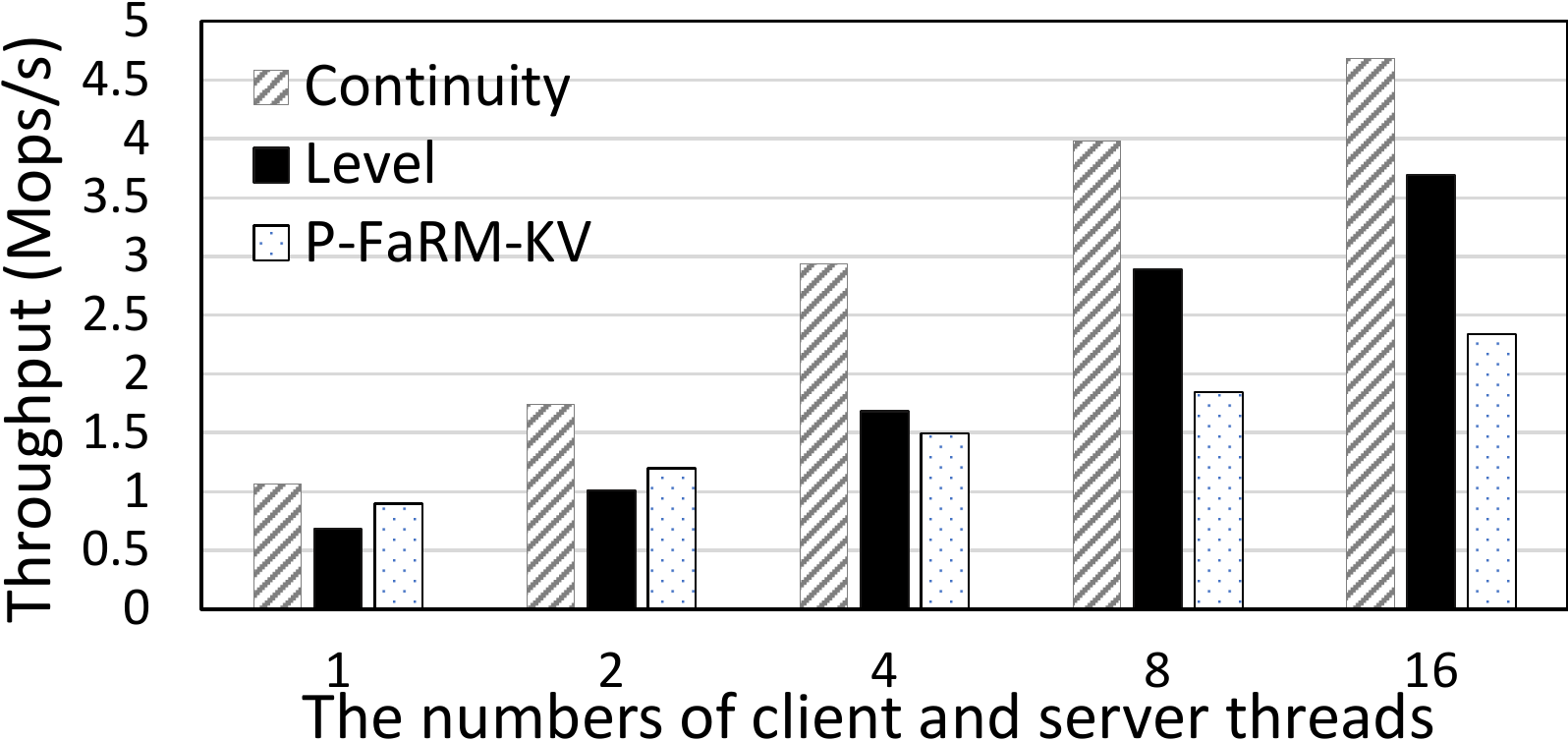}
		\setlength{\abovecaptionskip}{0.15cm}
		\setcaptionwidth{2.2in}
		\caption{\label{fig:throughput-workloadf}The throughput of YCSB-F (the read-modify-write workload).}
	\end{minipage}
	\vspace{-0.3cm}	
\end{figure*}

\begin{figure*}
	\begin{minipage}[t]{0.33\linewidth} 
		\centering
		\setlength{\abovecaptionskip}{0.1cm}
		\includegraphics[width=2.23in]{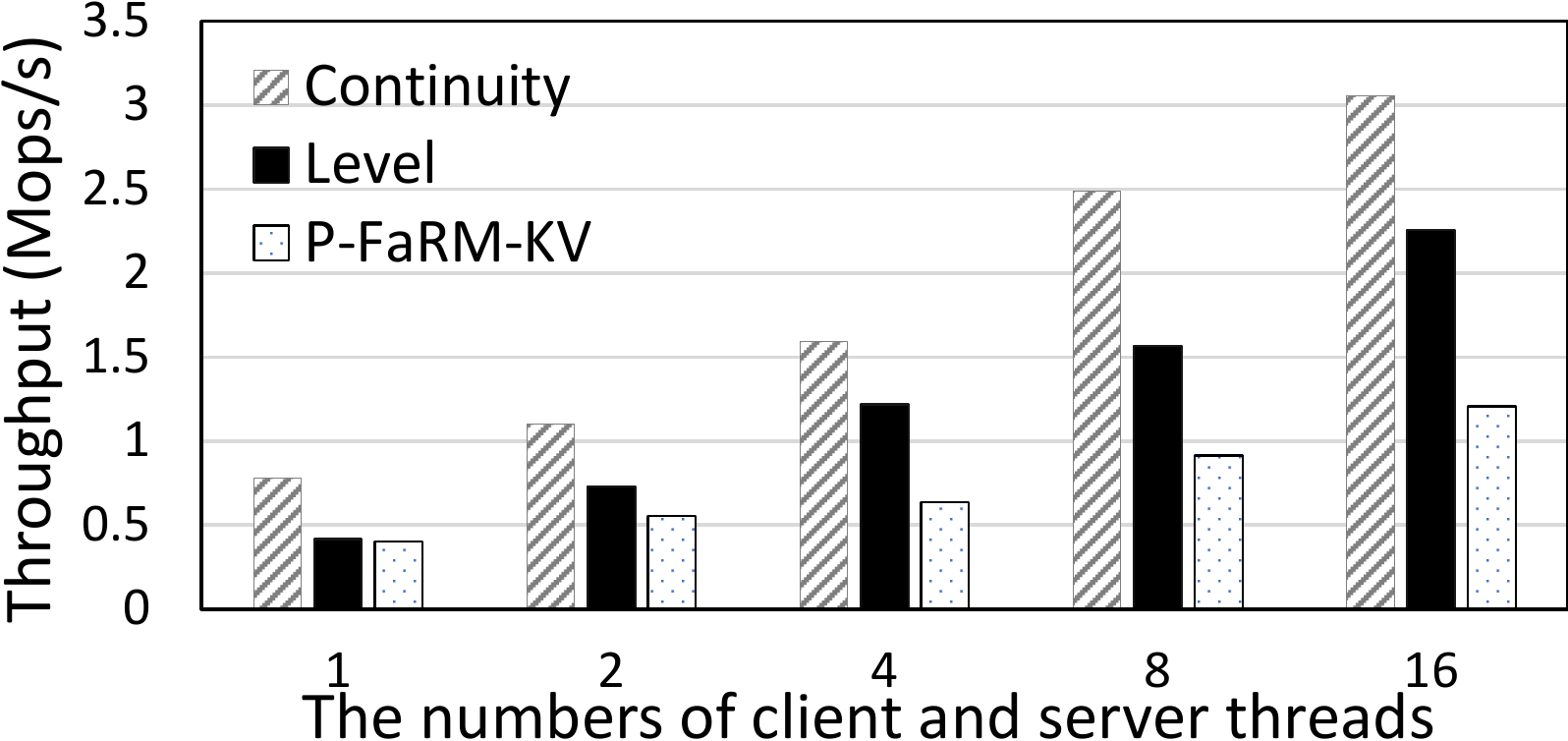}
		\setlength{\abovecaptionskip}{0.15cm}
		\setcaptionwidth{2.2in}
		\vspace{-0.4cm}
		\caption{\label{fig:throughput-update}The average throughput of 100\% updates.}
	\end{minipage}
    \begin{minipage}[t]{0.33\linewidth} 
    	\centering
    	\setlength{\abovecaptionskip}{0.1cm}
    	\includegraphics[width=2.23in]{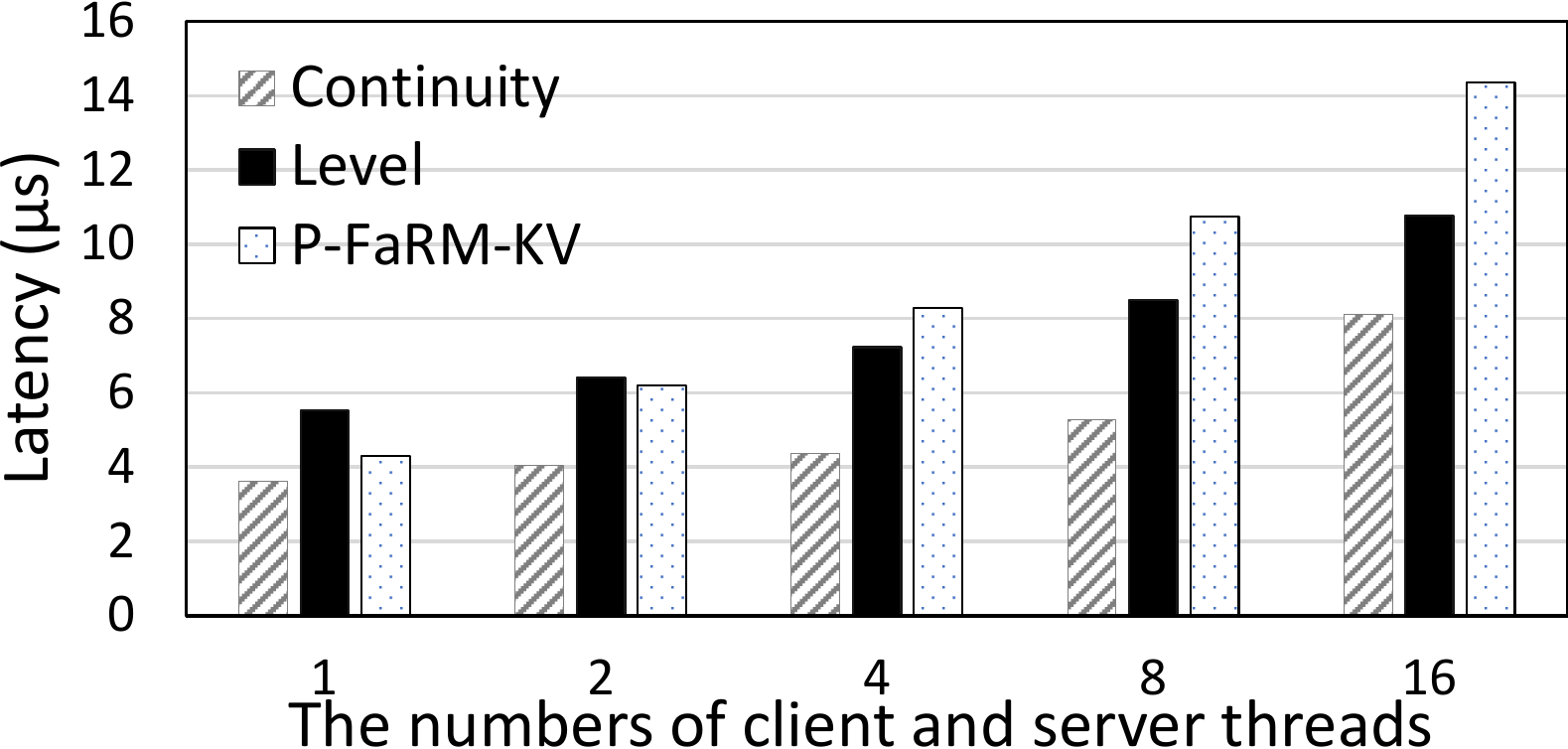}
    	\setlength{\abovecaptionskip}{0.15cm}
    	\setcaptionwidth{2.2in}
    	\caption{\label{fig:latency-workloada}The average latency of YCSB-A (the update-heavy workload).}
    \end{minipage}
    \begin{minipage}[t]{0.33\linewidth}
    	\centering
    	\setlength{\abovecaptionskip}{0.1cm}
    	\includegraphics[width=2.23in]{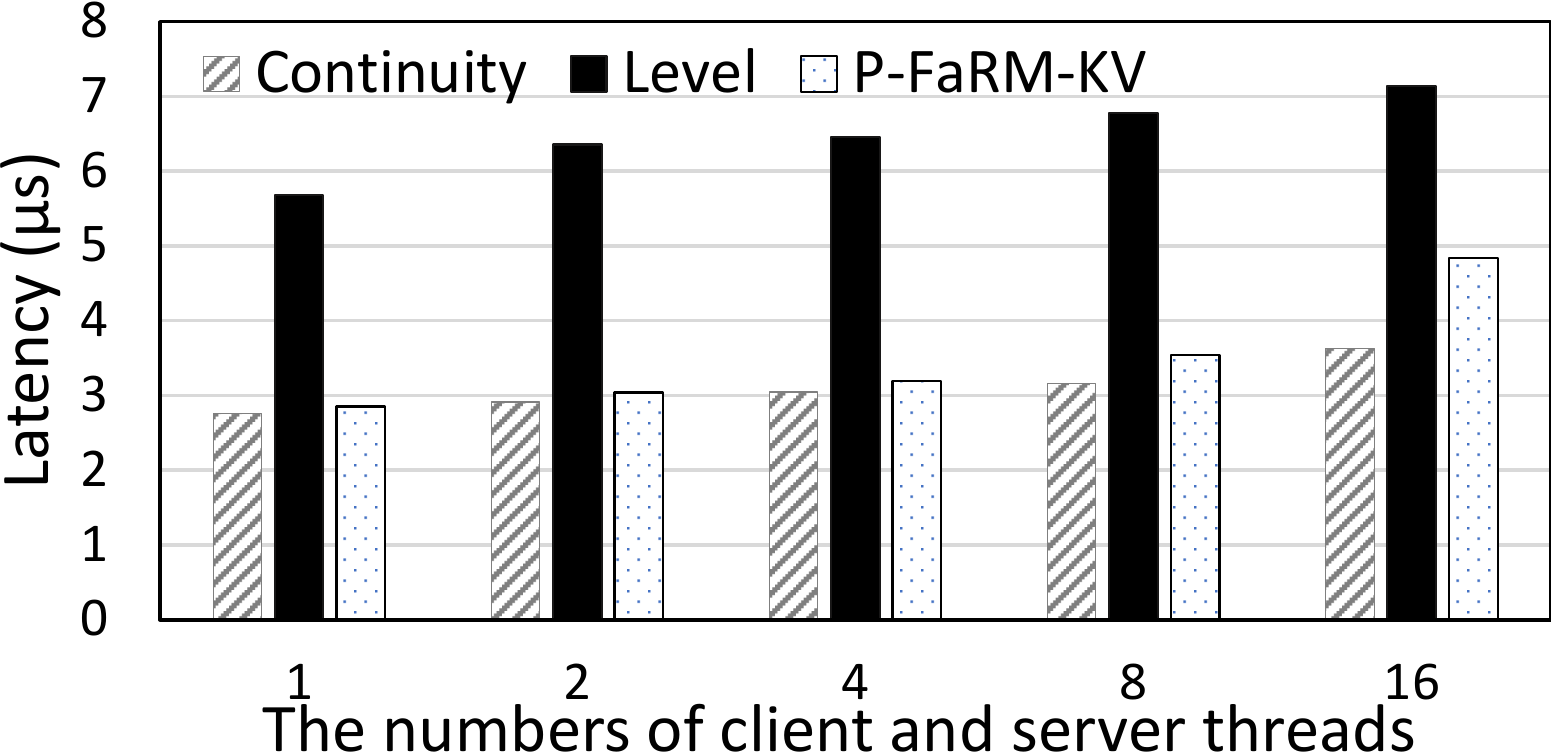}
    	\setlength{\abovecaptionskip}{0.15cm}
    	\setcaptionwidth{2.2in}
    	\caption{\label{fig:latency-workloadb}The average latency of YCSB-B (the read-mostly workload).}
    \end{minipage}
	\vspace{-0.3cm}	
\end{figure*}

\section{Performance Evaluation}\label{sec:evaluation}

\subsection{Experimental Setup}
Our experiments are performed on two Linux machines, each of which is equipped with a Mellanox ConnectX-$5$ Infiniband HCA, two $2.1$ GHz Intel Xeon $Gold 6230R$ CPUs, $192$GB DRAM and two $256$GB Optane DIMMs. We configure one Linux machine as a server and the other with up to $16$ threads as clients.

We generate our workloads via YCSB~\cite{cooper2010benchmarking}, which is a benchmark for key-value stores. Specifically, we use five workloads, i.e., YCSB-A, YCSB-B, YCSB-C, YCSB-D and YCSB-F. YCSB-A is the update-heavy workload, which consists of $50\%$ updates and $50\%$ reads. YCSB-B is the read-mostly workload, and consists of $95\%$ reads and $5\%$ updates. YCSB-C is the read-only workload. YCSB-D is the read-latest workload, where the most recently inserted records are the most popular. YCSB-F is the read-modify-write workload, and consists of $50\%$ reads and $50\%$ read-modify-writes, where a record will be read, then modified and written back. We also use individual operation microbenchmarks, i.e., insertion, deletion, update and search (YCSB-C). The evaluation does not include YCSB-E that exhibits range queries. For each workload, we execute $16$M key-value operations. Many schemes report that small-sized key-value pairs, e.g., a size of several or tens of bytes, dominate in production environment~\cite{chen2020flatstore,atikoglu2012workload,li2017kv,nishtala2013scaling}. Therefore, we follow the setting of level hashing, where the key size is $16$bytes, and the value size does not exceed $15$bytes~\cite{zuo2018write}.

Our proposed scheme is compared with two state-of-the-art hashing schemes, i.e., level hashing~\cite{zuo2018write} and P-FaRM-KV~\cite{dragojevic2014farm,lee2019recipe}. Level hashing is a PM-friendly hashing scheme. It provides single-node open-source code~\cite{levelcode}, and we add RDMA communication procedures to facilitate comparisons, i.e., using one-sided RDMA reads for remote read requests and RDMA write\_with\_immediate operations for remote write requests like our proposed continuity hashing. FaRM~\cite{dragojevic2014farm} is a general-purpose main memory distributed computing platform. On top of FaRM, FaRM-KV provides an RDMA-friendly hashing scheme for DRAM-based systems. We convert FaRM-KV into the PM counterpart with crash consistency guarantee, following the guidance of RECIPE~\cite{lee2019recipe}. Specifically, the original FaRM-KV supports lock-free RDMA reads, and uses logging to provide ACID transactions to ensure consistency, and thus RECIPE can convert the DRAM index into the PM version. The corresponding conversion guide we use is to insert cache line flush and mfence instructions after each store~\cite{lee2019recipe}. Note that even if we change the structure of FaRM-KV and add a bitmap to each bucket to support consistency within a bucket, FaRM-KV still needs to use logging when the updates occur across buckets for consistency. The evaluation does not include P-CLHT that is presented in RECIPE~\cite{lee2019recipe}, since P-CLHT uses linked lists to handle hash conflicts, which causes severe access amplification when using one-sided RDMA.

\subsection{Results and Analysis}
\subsubsection{Throughput}
Figures~\ref{fig:throughput-workloada} --~\ref{fig:throughput-workloadf} respectively show the average throughputs with various workloads for our continuity hashing, level hashing and P-FaRM-KV. The numbers of client and server threads vary from $1$ to $16$. The three hashing schemes use fine-grained locking for each slot to support concurrent access. For YCSB-A workload ($50\%$ updates, $50\%$ reads), the continuity hashing achieves $1.45X$ and $2.24X$ throughput improvements, compared with level hashing and P-FaRM-KV, as shown in Figure~\ref{fig:throughput-workloada}. The continuity hashing outperforms the PM-friendly level hashing, since querying data in the level hashing requires multiple one-sided RDMA round-trips. Moreover, continuity hashing also significantly outperforms the RDMA-friendly P-FaRM-KV, since the P-FaRM-KV fails to optimize PM writes and employs the expensive logging to guarantee consistency.

Figures~\ref{fig:throughput-workloadc} and~\ref{fig:throughput-workloadc-negative} respectively show the average throughputs of positive and negative searches. A \emph{positive search} means that the requested key-value item can be found in the hash table, while a \emph{negative search} is not. We observe that for level hashing, the average throughput of negative searches significantly decreases compared with that of positive searches, since for each negative search, the level hashing needs to issue four RDMA reads to query all the standby positions. Compared with level hashing, the continuity hashing and P-FaRM-KV achieve higher throughputs, i.e., $1.74X$ and $1.5X$ improvements for positive searches, and $3.07X$ and $2.46X$ improvements for negative searches. Because both continuity hashing and P-FaRM-KV are optimized for RDMA reads, and each query only needs nearly one RDMA operation. Figure~\ref{fig:throughput-update} shows the throughput of update-only workload. Continuity hashing and level hashing also achieve $2.43X$ and $1.66X$ throughput improvements compared with P-FaRM-KV, since both continuity hashing and level hashing optimize the high-overhead consistency mechanism for PM writes.

One read-modify-write operation in YCSB-F contains two access operations, i.e., a read and a write. As shown in Figure~\ref{fig:throughput-workloadf}, the average throughputs of YCSB-F for the three hashing schemes are lower than those of other YCSB workloads. However, our continuity hashing still achieves the highest throughput (i.e., $1.45X$ and $1.85X$ improvements) compared with the two state-of-the-art hashing schemes, i.e., level hashing and P-FaRM-KV.

\subsubsection{Latency}\label{sec:evaluation-Latency}
\begin{figure*}
	\begin{minipage}[t]{0.33\linewidth}
		\centering
		\setlength{\abovecaptionskip}{0.1cm}
		\includegraphics[width=2.23in]{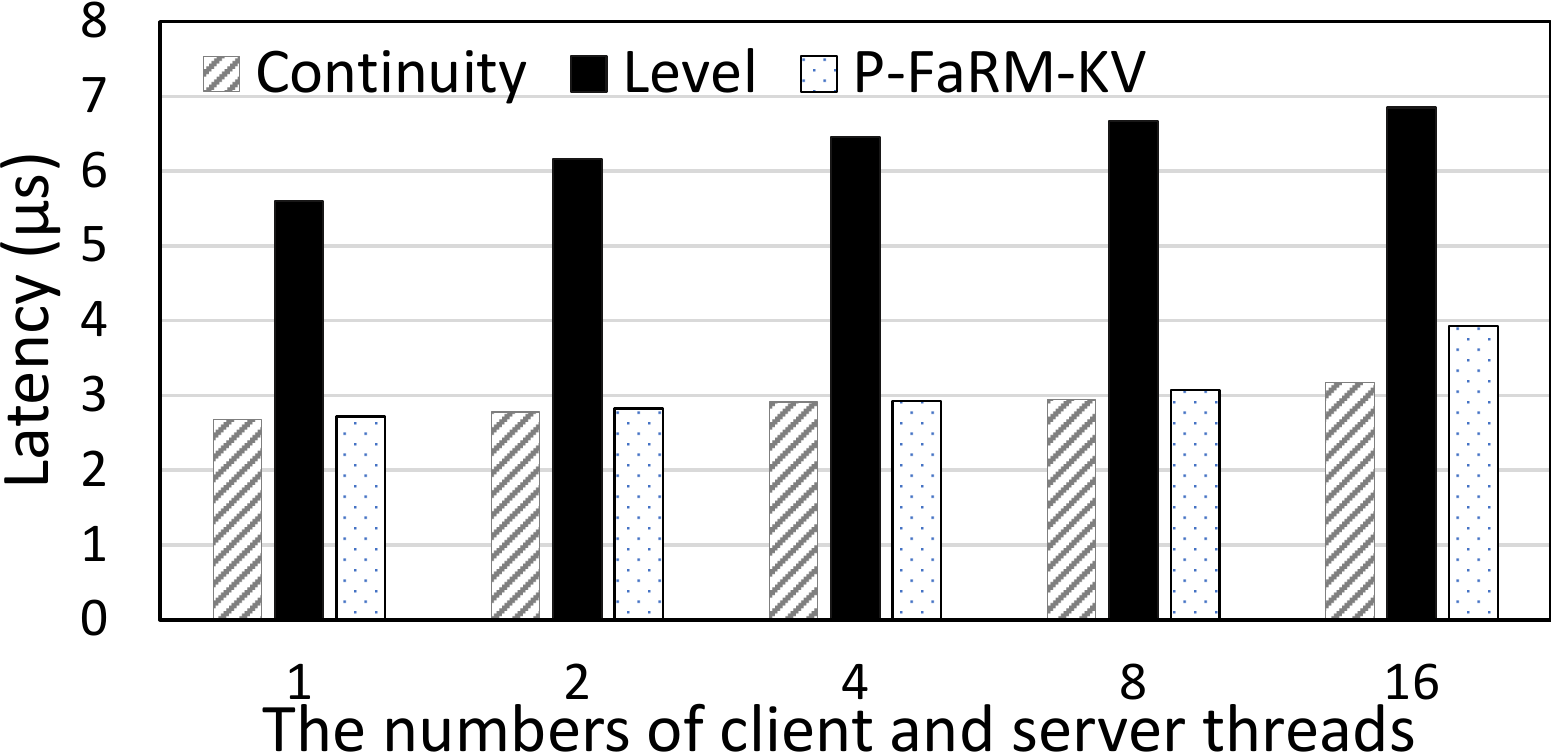}
		\setlength{\abovecaptionskip}{0.15cm}
		\setcaptionwidth{2.2in}
		\caption{\label{fig:latency-workloadc-positive}The latency of YCSB-C (the read-only workload, positive searches).}
	\end{minipage}
	\begin{minipage}[t]{0.33\linewidth}
		\centering
		\setlength{\abovecaptionskip}{0.1cm}
		\includegraphics[width=2.23in]{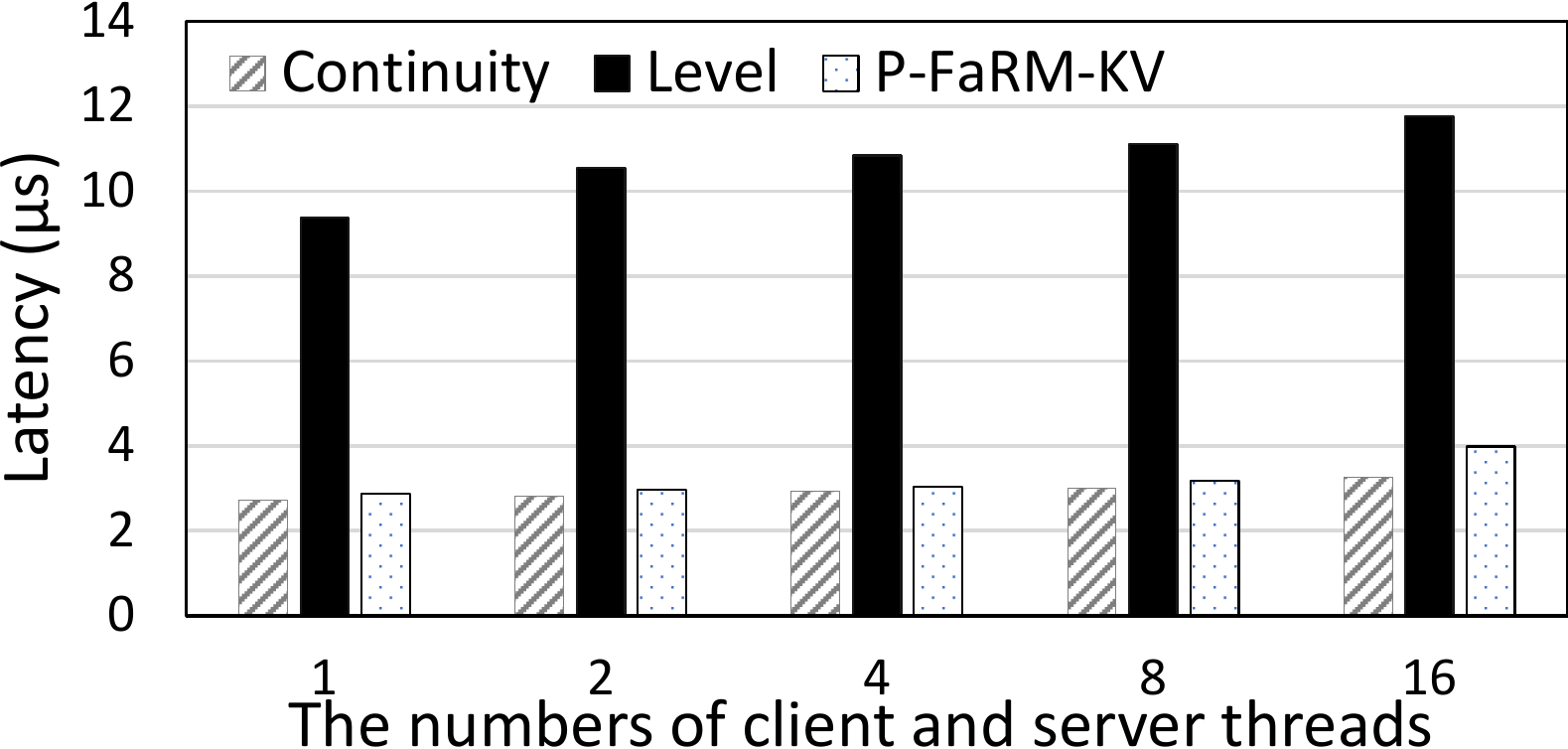}
		\setlength{\abovecaptionskip}{0.15cm}
		\setcaptionwidth{2.2in}
		\vspace{-0.4cm}	
		\caption{\label{fig:latency--workloadc-negative}The average latency of negative searches.}
	\end{minipage}
	\begin{minipage}[t]{0.33\linewidth} 
		\centering
		\setlength{\abovecaptionskip}{0.1cm}
		\includegraphics[width=2.23in]{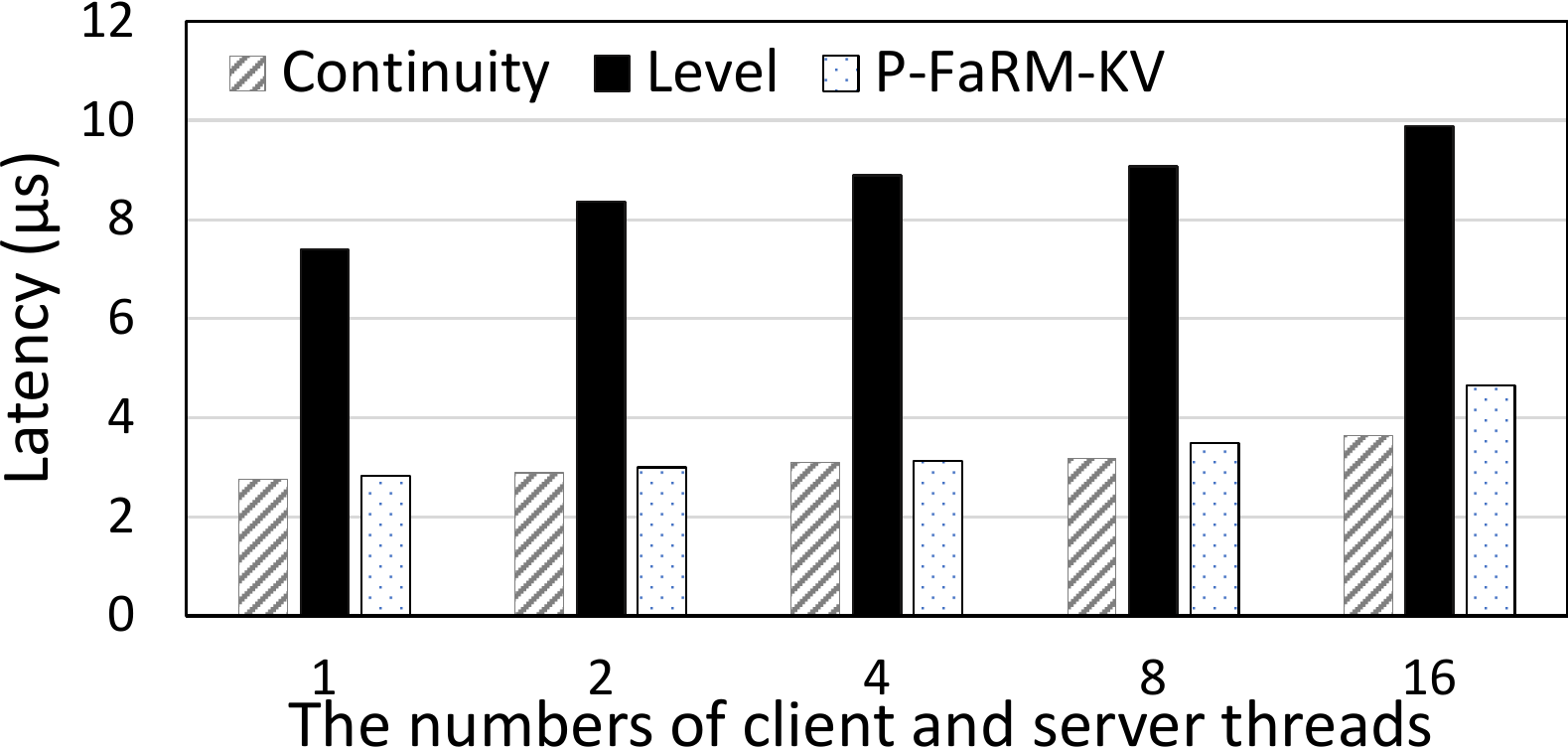}
		\setlength{\abovecaptionskip}{0.15cm}
		\setcaptionwidth{2.2in}
		\caption{\label{fig:latency-workloadd}The latency of YCSB-D (the read-latest workload).}
	\end{minipage}
	\vspace{-0.3cm}	
\end{figure*}

We record the execution time of each individual operation as latency. Figures~\ref{fig:latency-workloadb} --~\ref{fig:latency-workloadd} show the average latencies of four types of read-dominated workloads, i.e., YCSB-B, YCSB-C (positive searches), negative searches and YCSB-D. We observe that compared with level hashing and P-FaRM-KV, the continuity hashing reduces the latency by an average of $61\%$ and $9\%$. Note that for level hashing, the latency of YCSB-D is much higher than that of YCSB-B and YCSB-C. YCSB-D is the read-latest workload, and hence the data to be read are more likely to be located in the hash bucket with lower access priority among the four alternative buckets, thus resulting in the potential multiple RDMA read round-trips.

\begin{figure*}
	\begin{minipage}[t]{0.33\linewidth}
		\centering
		\setlength{\abovecaptionskip}{0.1cm}
		\includegraphics[width=2.23in]{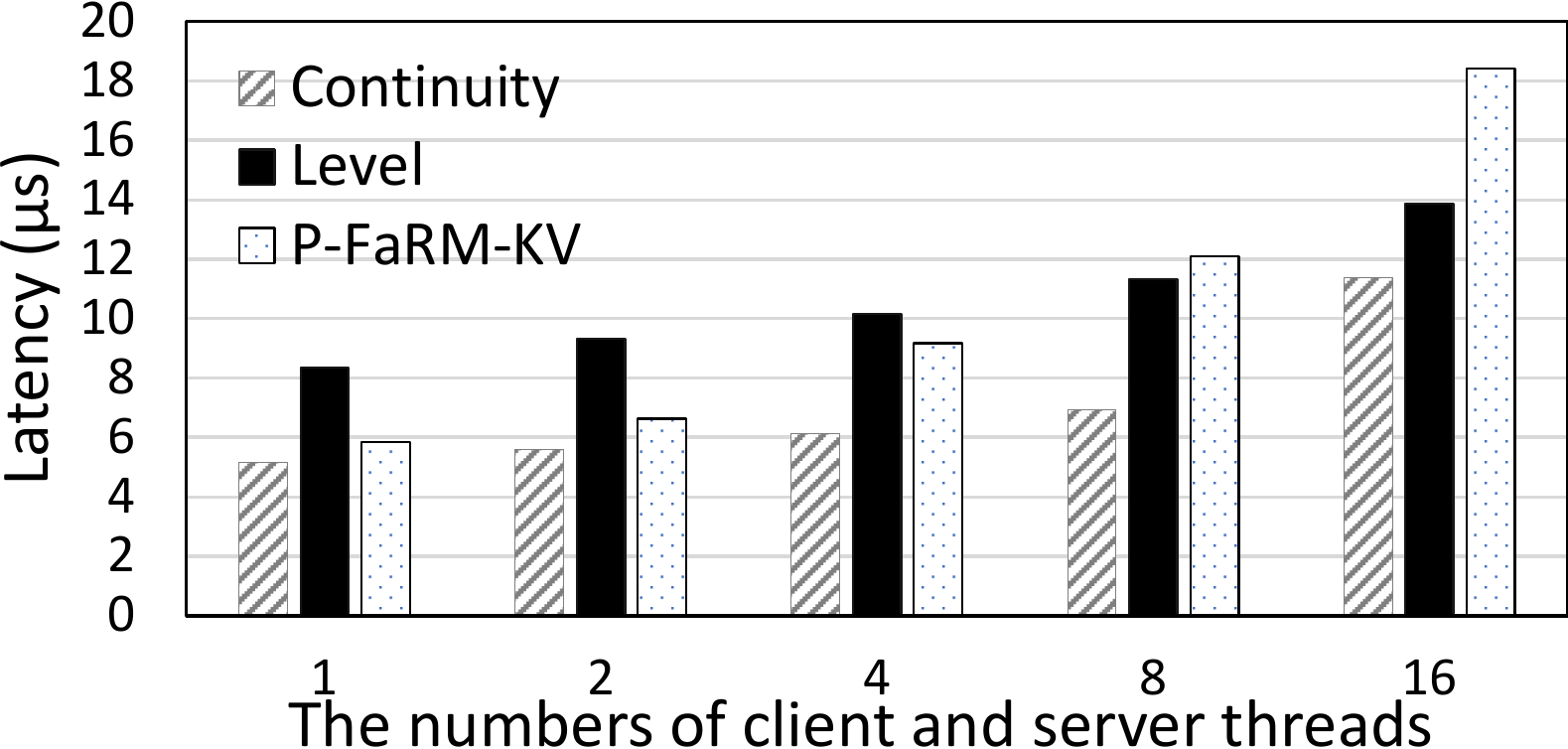}
		\setlength{\abovecaptionskip}{0.15cm}
		\setcaptionwidth{2.2in}
		\caption{\label{fig:latency-workloadf}The average latency of YCSB-F (the read-modify-write workload).}
	\end{minipage}
	\begin{minipage}[t]{0.33\linewidth}
		\centering
		\setlength{\abovecaptionskip}{0.1cm}
		\includegraphics[width=2.23in]{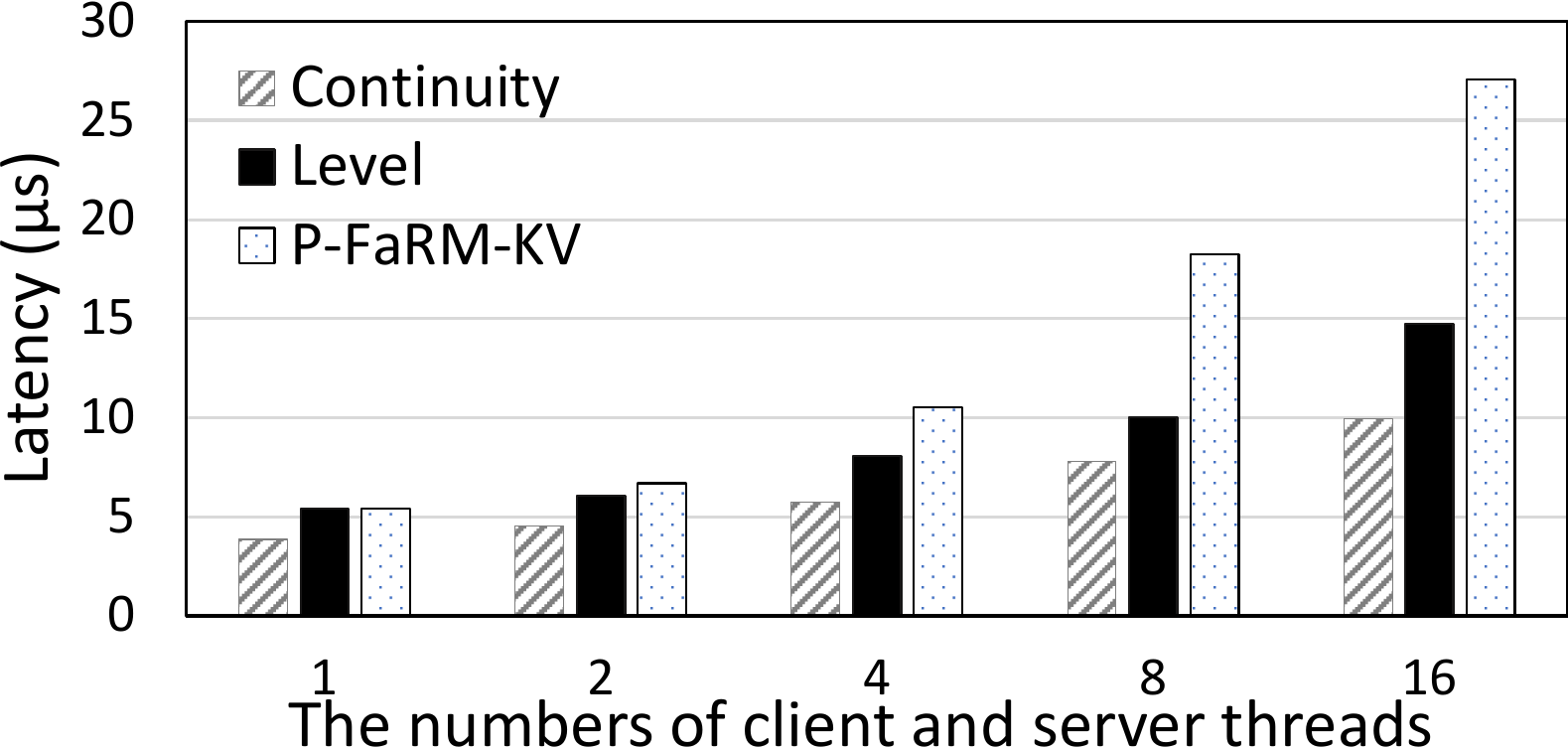}
		\setlength{\abovecaptionskip}{0.15cm}
		\setcaptionwidth{2.2in}
		\caption{\label{fig:latency-update}The average latency of 100\% updates.}
	\end{minipage}
	\begin{minipage}[t]{0.33\linewidth}
		\centering
		\setlength{\abovecaptionskip}{0.1cm}
		\includegraphics[width=2.23in]{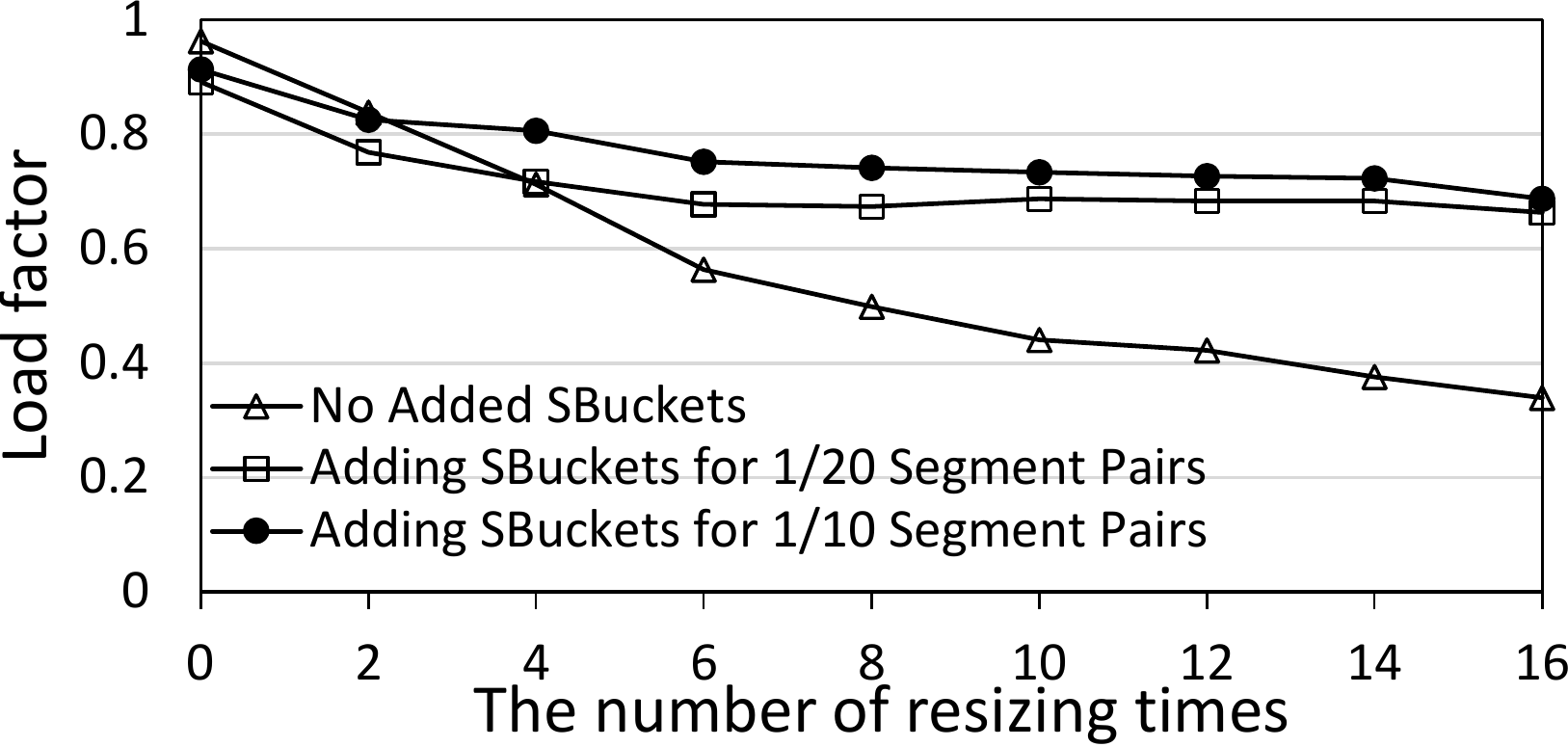}
		\setlength{\abovecaptionskip}{0.15cm}
		\setcaptionwidth{2.2in}
		\caption{\label{fig:load-factor}The load factors of continuity hashing with different schemes.}
	\end{minipage}
	\vspace{-0.3cm}	
\end{figure*}

Figures~\ref{fig:latency-workloada} and~\ref{fig:latency-workloadf} show the average latencies of YCSB-A and YCSB-F, which contains a number of write operations. The average latency of writes increases with the number of threads due to locking mechanism.  Compared with level hashing and P-FaRM-KV, the continuity hashing reduces the latency by an average of $34\%$ and $37\%$, i.e., speeding up the operations in YCSB-A and YCSB-F workloads by $1.5X$ and $1.6X$ on average. As shown in Figure~\ref{fig:latency-update}, for the latencies of PM update operations, our continuity hashing respectively achieves $1.39X$ and $2.14X$ performance improvements on average, compared with the PM-write-friendly level hashing and the RDMA-friendly P-FaRM-KV. We have optimized the insertion operations of P-FaRM-KV to reduce PM writes and latency by replacing the iteratively displacing key-value pairs in the original scheme with at most one movement. However, the logging consistency mechanism causes double PM writes and hurts the write performance of P-FaRM-KV. Continuity hashing is optimized for PM writes and leverages a log-free consistency scheme. The results further demonstrate our performance advantages over the two state-of-the-art schemes.

\subsubsection{The Number of PM Writes}
We evaluate the number of PM writes by counting the number of cache line flush instructions for each write operation. Table~\ref{table01} shows the results for insertion, update and deletion operations. Specifically, in our continuity hashing, each insertion and update needs to sequentially write the key-value pair in an empty slot and modify the associated bit in the indicator from $0$ to $1$, thus including $2$ PM writes. Furthermore, a deletion operation only needs $1$ PM write that modifies the associated bit from $1$ to $0$. In  the level hashing, the numbers of PM writes for an insertion or an update is affected by the load factor of hash table. When the load rate increases, the probability of using logging to ensure crash consistency will also increase, which causes more PM write operations. The number of PM writes for an insertion and an update are $2 - 2.01$ and $2 - 5$, respectively. The deletion operation in the level hashing is similar to that in the continuity hashing. P-FaRM-KV uses logging for each write operation, which incurs the highest number of PM writes among the three hashing schemes.

\begin{table}[!hbp]
	\renewcommand\arraystretch{1.19}
	\vspace{-8px}
	\caption{\label{table01} The number of PM writes with different operations.}
	\vspace{0.3cm}	
	\centering
	\arrayrulewidth0.8pt
	\small
	\vspace{1px}
	\begin{tabular}{|c|c|c|c|}
		\hline \tabincell{c}{}&Insertion&Update&Deletion\\
		\hline
		\hline Continuity&2&2&1\\
		\hline Level&2 -- 2.01&2 -- 5&1\\
		\hline P-FaRM-KV&5&5&5\\
		\hline
	\end{tabular}
\end{table}

\subsubsection{Maximum Load Factor}
We evaluate the optimization scheme proposed in Section~\ref{subsec:Space Utilization}, which dynamically adds SBuckets for a small percentage of segment pairs to improve load factors. The evaluation uses YCSB-A workload. In Figure~\ref{fig:load-factor}, we evaluate the load factors of continuity hashing and show the effectiveness of the added SBuckets in terms of space utilization. The initial hash table contains $20$ buckets (i.e., $80$ slots). Each resizing expands the hash table to twice the current capacity. The x-axis represents the number of resizing times. The y-axis represents the load factor each time the next resizing operation will be triggered. We observe that as the number of resizing increases, the load factors of our original solution without added SBuckets gradually decrease. The optimization schemes with the added Sbuckets for $1/20$ and $1/10$ segment pairs achieve the load factors of about $70\%$, which is acceptable due to the large capacity of the available PM products~\cite{izraelevitz2019basic}. We further evaluate the throughput of continuity hashing with different optional schemes. The results are omitted here due to space limit. The results show that adding SBuckets for $1/10$ and $1/20$ segment pairs only slightly reduces the throughput of YCSB-A by $4\%$ -- $5\%$.

\subsection{Discussion}\label{sec:discussion}

The continuity hashing supports to handle key-value pairs of different sizes via storing the key and a fat pointer for a large key-value pair in the bucket. The fat pointer indicates the position and size of the corresponding value. Storing keys and values separately is one way to deal with large-sized key-value pairs, although this paper focuses on the small key-value pairs that dominate in production environments.

\section{Related Work}\label{sec:related work}

\textbf{RDMA-based Hashing Schemes.} In order to exploit the fast RDMA networking and improve system performance with low CPU overhead, existing schemes propose RDMA-based hashing index structures~\cite{wei2015fast,dragojevic2014farm,mitchell2013using}. Pilaf~\cite{mitchell2013using} adopts $3$-way cuckoo hashing with a novel self-verifying hash table structure. Each probe for looking up a key requires two one-sided RDMA reads (i.e., one for the hash table entry and the other for the actual key-value pair). The proposed self-verifying data structure is used to detect read-write races, where each hash table entry requires two $64$-bit checksums. HERD~\cite{kalia2014using} focuses on fully exploring RDMA features to build high-performance key-value stores. This work uses RDMA write operations over unreliable connection (UC) mode and RDMA send operations over unreliable datagram (UD) mode to handle put and get requests. DrTM-KV~\cite{wei2015fast,chen2016fast,wei2018deconstructing} proposes the HTM/RDMA-friendly cluster hashing based on the traditional chained hashing, which leverages clustering and location-based caches to reduce the number of one-sided RDMA operations. FaRM~\cite{dragojevic2014farm,dragojevic2015no,shamis2019fast} proposes a hash table that is a variant of hopscotch hashing with chaining and associativity. FaRM-KV enables high space efficiency and remote query with a small number of one-sided RDMA reads. However, these RDMA-based hashing schemes fail to address the problems of the limited write endurance and crash consistency on PM. Unlike them, our proposed continuity hashing is optimized for both RDMA and PM.

\textbf{PM-based Hashing Schemes.} In order to leverage the non-volatile and byte-addressable properties of PM and improve index reliability and efficiency, existing schemes propose PM-based hashing index structures~\cite{lee2019recipe,nam2019write,zuo2018write,zuo2017write,debnath2016PFHT}. Level hashing~\cite{zuo2018write} is a write-optimized two-level hashing scheme for PM, providing cost-efficient resizing and low-overhead consistency guarantee. CCEH~\cite{nam2019write} is a PM-friendly three-level dynamic hashing. It makes effective use of cache lines and guarantees crash consistency without explicit logging. RECIPE~\cite{lee2019recipe} proposes to convert DRAM indexes that meets the specified conditions into their PM counterparts with crash consistency guarantee. PFHT~\cite{debnath2016PFHT} is a variant of cuckoo hashing for reducing writes to phase change memory. Each key-value item can be stored in two buckets in the main table, and a supplementary stash stores any overflow key-value pairs to improve the load factor. Path hashing~\cite{zuo2017write} is a write-friendly hashing scheme for PM. The buckets in the path hashing are logically organized as an inverted binary tree. The leaf nodes of the binary tree are addressable, while the non-leaf nodes in the same path are used as a stash. However, when these PM-based hashing schemes are applied to RDMA environments, a remote read request usually results in the need for multiple one-sided RDMA round-trips. Unlike them, our proposed continuity hashing is a coalescing design for RDMA and PM, which aims to reduce RDMA access amplification and PM writes, as well as ensuring crash consistency.

\section{Conclusion}\label{sec:conclusion}
Designing a high-performance hash structure is important for memory systems based on coalescing RDMA and PM. In order to address the problems of RDMA Access Amplification and High-Overhead PM Consistency, we propose the continuity hashing, a coalescing hashing solution for both RDMA and PM. The continuity hashing supports efficient remote read via a single one-sided RDMA operation and log-free consistency guarantee for all the write operations on PM. The evaluation demonstrates that our proposed continuity hashing achieves the high throughput, the low latency as well as the small number of PM writes, while obtaining acceptable load factors.

{\bibliographystyle{./IEEEtran}
\bibliography{references}}

\end{document}